\documentclass{aastex6}
\pdfoutput=1
\newcommand{\kms}{\ensuremath{\rm{km\,s^{-1}}}}
\usepackage[utf8]{inputenc}
\begin{document}
\shorttitle{SMC ISM Element Depletions}
\shortauthors{Jenkins \& Wallerstein}
\title{INTERSTELLAR GAS-PHASE ELEMENT DEPLETIONS IN THE SMALL 
MAGELLANIC CLOUD:\\
A GUIDE TO CORRECTING FOR DUST IN QSO ABSORPTION LINE 
SYSTEMS\footnote{Based on observations with the NASA/ESA Hubble Space 
Telescope and additional data obtained from the Data Archive at the Space 
Telescope Science Institute, which is operated by the Associations of Universities for 
Research in Astronomy, Incorporated, under NASA contract NAS5-26555. These 
observations are associated with program nr. 13778}\footnote{\copyright 2017. The American
Astronomical Society. All rights reserved.}}

\author{Edward B. Jenkins}
\affil{Princeton University Observatory, Princeton, NJ, 08544-1001}
\email{ebj@astro.princeton.edu}
\and
\author{George Wallerstein}
\affil{University of Washington, Seattle, Dept. of Astronomy,
Seattle, WA 98195-1580}
\email{walleg@u.washington.edu}

\begin{abstract}
We present data on the gas phase abundances for 9 different elements in the 
interstellar medium of the Small Magellanic Cloud (SMC), based on the strengths of 
ultraviolet absorption features over relevant velocities in the spectra of 18 stars 
within the SMC.  From this information and the total abundances defined by the 
element fractions in young stars in the SMC, we construct a general interpretation 
on how these elements condense into solid form onto dust grains.  As a group, the 
elements Si, S, Cr, Fe, Ni, and Zn exhibit depletion sequences similar to those in the 
local part of our Galaxy defined by Jenkins (2009).  The elements Mg and Ti deplete 
less rapidly in the SMC than in the Milky Way, and Mn depletes more rapidly.  We 
speculate that these differences might be explained by the different chemical 
affinities to different existing grain substrates.  For instance, there is evidence that 
the mass fractions of polycyclic aromatic hydrocarbons (PAHs) in the SMC are 
significantly lower than those in the Milky Way.  We propose that the depletion 
sequences that we observed for the SMC may provide a better model for 
interpreting the element abundances in low metallicity Damped Lyman Alpha (DLA) 
and sub-DLA absorption systems that are recorded in the spectra of distant quasars 
and gamma ray burst afterglows.
\end{abstract}


\section{INTRODUCTION}\label{sec:intro}

When combined with theories of stellar evolution and nucleosynthesis, 
measurements of the abundances of elements in different parts of any galactic 
system reveal much about how it formed and evolved, how and when it exchanged 
gas with the intergalactic medium (both infall and outflow), and how its stellar 
populations changed with time. Stars of different masses, initial compositions, and 
age produce their own distinct imprint of element production (Wheeler et al. 1989 ; 
Timmes et al. 1995 ; McWilliam 1997 ; Chiappini et al. 1999 ; Matteucci 2003 ; 
Kobayashi et al. 2006). For our own Galaxy, objects are close enough that we can 
examine in great detail how different element groups, such as $\alpha$, Fe-peak, 
neutron-capture or cosmic ray spallation products, change from one location (or 
star) to the next.  From theories of nucleosynthesis, we know that even-Z elements 
of intermediate mass, such as Mg, S, Si, Ca, and probably Ti arise primarily from 
fundamental reactions in massive stars and core-collapse SNe, while the odd-Z 
elements such as P, Na, Al, and K depend on having a neutron excess and thus are 
driven by the initial metallicities of the stars (Suess \& Urey 1956 ; Burbidge et al. 
1957 ; Cameron 1957). Type~1a supernovae are mostly responsible for the 
production of the Fe-group elements (V, Cr, Mn, Fe, Co, and Ni). In the studies of 
individual stars, we are able to trace the mix of these groups as a function of stellar 
ages (conventionally traced by their relative abundances of iron [Fe/H]), or by their 
memberships in dynamically distinct populations (thin disk, thick disk, bulge, etc.).

For other galaxies, we do not have access to the tremendous level of detail that we 
have from nearby.  Nevertheless, the global values of some element abundances and 
their gradients across the surfaces of the galaxies can still be studied from the 
spectroscopy of emission lines from H~II regions
(Thuan et al. 1995 ; Izotov \& Thuan 1999 ; Chen et al. 2005 ; Christensen et al. 2005 
; Ellison et al. 2005 ; Schulte-Ladbeck et al. 2005 ; Péroux et al. 2011 ; 2012 ; 2014).  
This information is useful for gaining a better understanding of the changes caused 
by internal processes, mass loss, and mergers from one galaxy to the next.  However, 
to go to even greater distances and explore the elemental makeup of galaxies during 
early times in the history of our universe, i.e., at redshifts $z\gtrsim 2.5$, it becomes 
difficult to obtain much spectroscopic detail from the light that is emitted.  Virtually 
all of our knowledge on element abundances arises from studies of UV absorption 
lines, redshifted to visible wavelengths, that are seen in the spectra of background 
quasars or the afterglows of gamma ray bursts.  

From UV absorption-line studies of interstellar material in our own Galaxy, we know 
that the gas-phase abundances of different elements are depleted by condensation 
into solid form within interstellar dust grains.\footnote{Contrary to some early 
misconceptions, the elements O, S, and Zn are \underbar{\it not\/} undepleted.  
Only N seems to be mostly undepleted.}   We have known for some time that the 
strengths of these depletions vary strongly from one element to the next and from 
one sight line to another (Habing 1969 ; Wallerstein \& Goldsmith 1974 ; Morton 
1975 ; Jenkins et al. 1986 ; Savage \& Sembach 1996 ; Dwek 2016).  Researchers 
who study quasar absorption-line systems, such as Damped Lyman Alpha systems 
(DLAs having $\log N({\rm H~I})/{\rm cm}^{-2}\gtrsim 20.3$) or sub-DLAs ($18 
\lesssim \log N({\rm H~I})/{\rm cm}^{-2} \lesssim 20.3$) have had to use what 
we have learned from the depletion patterns in the Milky Way to make corrections 
that will help to define the pattern of the true, intrinsic abundances of any system 
(Pettini et al. 1994 ; Lu et al. 1996 ; Kulkarni et al. 1997 ; Pettini et al. 1997 ; Pettini 
et al. 1999 ; Prochaska \& Wolfe 1999 ; Pettini et al. 2000 ; Hou et al. 2001 ; Ledoux 
et al. 2002 ; Prochaska \& Wolfe 2002 ; Calura et al. 2003 ; Prochaska et al. 2003 ; 
Dessauges-Zavadsky et al. 2004 ; Vladilo 2004 ; Lopez et al. 2005 ; Dessauges-
Zavadsky et al. 2006 ; Rodríguez et al. 2006 ; Levshakov et al. 2009 ; Meiring et al. 
2009 ; Cooke et al. 2011 ; Rafelski et al. 2012 ; Som et al. 2013 ; Fox et al. 2014 ; 
Kulkarni et al. 2015 ; Prochaska et al. 2015 ; Som et al. 2015 ; Guber \& Richter 2016 
; Morrison et al. 2016 ; Quiret et al. 2016 ; Wiseman et al. 2016).  Some early 
attempts to characterize the depletions of certain elements in order to correct for 
them were devised by Savaglio (2001), Vladilo (2002) and Prochaska \& Wolfe 
(2002).

In a comprehensive review of the gas-phase abundances of 17 different elements in 
the interstellar medium in our local region of the Milky Way, Jenkins (2009) 
presented an analysis that showed how strongly each element depletes into solid 
form (dust) as the overall levels of depletions for the other elements change from 
one sight line to the next.  A remarkable finding from this investigation was that the 
logarithms of the depletion factors of different elements tracked each other in linear 
fashions, but at different rates.  This unified picture of depletion greatly simplified 
our understanding of how elements disappear from the gas phase and bind into 
solid form.  Going further, we were able to learn from these results that the 
proportions of different the atomic constituents of grains change as the overall 
severity of depletions changed (Jenkins 2009 ; 2013). 

\section{MOTIVATION}\label{sec:motivation}

In principle, one might suppose that it should be a simple matter to use our 
knowledge of how depletions behave in our Galaxy to make corrections for such 
processes in other environments.  Indeed, from the lack of any better choice many 
investigations invoked this method to determine the total element abundances in 
distant systems, most of which had metallicities of order 1/300 to 1/3 solar 
(Dessauges-Zavadsky et al. 2006 ; Dessauges-Zavadsky et al. 2007 ; Rafelski et al. 
2014 ; Quiret et al. 2016).  However, evidence has accumulated that shows that the 
depletion patterns probably change when the basic element abundances or 
production sequences are different.   For example, in the Milky Way Si and Fe 
deplete at about the same rate, and the gas-phase logarithmic abundance ratio 
relative to the solar one [Si/Fe]$_{\rm gas}\approx 0.8$.  However,  Wolfe et al. 
(2005) showed that without corrections for depletion [Si/Fe]$_{\rm gas}$ starts at 
about 0.3 for DLAs with $[{\rm Si/H}]_{\rm gas}\lesssim -1.0$, and increases only 
slightly as the metallicities approach that of our Galaxy.  After we correct for 
depletion, we would say that the intrinsic (i.e., gas plus dust) $[{\rm Si/Fe}]\approx 
-0.5$ for these systems, a trend that is contrary to $[{\rm Si/Fe}]\approx +0.5$ for 
stars with $[{\rm Fe/H]}<-1.5$ in the Milky Way (Timmes et al. 1995).  Likewise, 
Ledoux et al. (2002) stated that ``The correlation between [Mn/Fe] and [Zn/Fe] ... 
cannot be accounted for by any dust depletion sequence: it implies either variations 
of the intrinsic Mn abundance relative to Fe from $-0.3$ to +0.1~dex and/or a 
relation between depletion level and metallicity.''  They also stated that ``The 
variations of [Ti/Fe] vs. [Zn/Fe] cannot be fitted by a single dust depletion sequence 
either.''  These abnormalities may be explained by chemical considerations in pre-
existing solids: for instance, Lodders (2003) presented examples where the 
condensation of the elements Ni and Ge depend on the presence of a host element 
Fe to create an alloy.    Likewise, the formation of refractory compounds that contain 
Zn and Mn, such as ${\rm Zn_2SiO_4}$, ${\rm ZnSiO_3}$, or ${\rm Mn_2SiO_4}$, are 
aided by host minerals such as forsterite and enstatite.  In some galaxy 
environments where the ratio of $\alpha$-group to Fe-peak elements differs from 
that of our own, or where the previous buildup of some element groups diverged 
from that of the Milky Way, certain elements may have had more or less than their 
respective host compounds, thus altering the depletion rates in a way that is difficult 
to predict.

From the preceding discussion, we can see a clear need to investigate the depletion 
sequences in a low-metallicity system, much as Jenkins (2009) had done with solar 
metallicity gas in the local part of our Galaxy.  The Small Magellanic Cloud (SMC) is a 
nearby dwarf irregular galaxy that presents an excellent opportunity to perform 
such a definition.  It has a considerably lower metallicity than that of the Milky Way, 
and its chemical evolution history might possibly present a better match to more 
distant galaxies with metallicities $[{\rm M/H}] \sim -1$ that are typically 
represented by DLAs at redshifts $z \lesssim 3.5$ (Rafelski et al. 2012).  Moreover, 
most of its stars indicate foreground extinction curves that differ from those of stars 
in the Milky Way (Hutchings 1982 ; Bromage \& Nandy 1983 ; Prevot et al. 1984 ; 
Gordon \& Clayton 1998 ; Gordon et al. 2003 ; Cartledge et al. 2005 ; Maíz Apellániz 
\& Rubio 2012 ; Hagen et al. 2016), which suggests deviations in the distributions of 
dust grain sizes and/or compositions (Draine \& Lee 1984 ; Boulanger et al. 1994 ; 
Weingartner \& Draine 2001 ; Zubko et al. 2004 ; Zonca et al. 2015).  This difference 
could be relevant to studies of abundances in DLAs, since their extinction curves are 
similar to that of the SMC (Murphy \& Bernet 2016).  As with our Galaxy, we can 
measure the abundances of elements in young stars, although with less accuracy.  
These stellar abundances can serve as a standard for the combined element 
abundances in both gas and dust.

There have been a number of studies of ISM abundances in the SMC that have 
already been carried out using data from spectrographs on the {\it Hubble Space 
Telescope\/} (HST) and the {\it Far Ultraviolet Spectroscopic Explorer\/} (FUSE)  
(Roth \& Blades 1997 ; Welty et al. 1997 ; Koenigsberger et al. 2001 ; Mallouris et al. 
2001 ; Welty et al. 2001 ; Sofia et al. 2006).  A dispute on Si abundances between 
Sofia et al. (2006) and Welty et al. (2001) for a single sight line, amounting to 
0.19$\,$dex, highlights the difficulties that have been encountered previously.  
Tchernyshyov et al. (2015) performed a more comprehensive survey that derived 
abundances of the elements Si, P, Cr, and Fe for many stars in the SMC using spectra 
recorded by the {\it Cosmic Origins Spectrograph\/} (COS) on the HST.  (We 
compare our results to theirs in Section~\ref{sec:T15}.)  Our approach differs from 
previous ones by obtaining medium resolution STIS echelle data with broad 
wavelength coverages that enabled us to cover many transitions of differing 
strengths for a significant collection of sight lines. 

\section{SELECTION OF TARGET STARS}\label{sec:selection}

Our investigation of element depletions made use of 18 sight lines toward stars in 
the SMC.  Table~\ref{tbl:target_stars} lists the stars used in this study.   Fourteen of 
the target stars were observed in our Cycle~22 observing program (46 orbits, 
Program ID = 13778, E.~Jenkins, PI) on the {\it Hubble Space Telescope\/} (HST).  
In selecting which stars to observe, we made use of the SMC interstellar titanium 
abundances reported by Welty \& Crowther (2010) as a guide for sampling a wide 
selection of relative depletions.  As a supplement to our observations, an additional 
four stars were observed for other programs, which produced suitable spectra that 
were publicly available in the {\it Mikulski Archive for Space Telescopes\/} (MAST) 
maintained by the Space Telescope Science Institute.  Nearly all of the stars had 
measurements of atomic and molecular hydrogen column densities reported by 
Welty et al. (2012) for gas only within the SMC.   For most stars, the combined 
column densities of hydrogen in both atomic and molecular form exceeded 
$10^{21}{\rm cm}^{-2}$; only 3 stars had lower values (see 
Table~\ref{tbl:col_dens_depl1} in Section~\ref{sec:depl_outcomes}).  For such 
high column densities, corrections for unseen ionization stages should be negligible 
(Vladilo et al. 2001).  Another important criterion in selecting stars was to insure 
that the projected rotational velocities $v\sin i > 50\,\kms$, so that stellar features 
would not create confusing continuum levels.

\section{OBSERVING STRATEGY AND DATA 
ANALYSIS}\label{sec:observing_strategy}

While it is usually desirable to obtain spectra at the highest possible wavelength 
resolution for analyzing interstellar absorptions, especially if the features are 
strong, we decided that a broad coverage in wavelength would register many 
different transitions of different strengths, which outweighed the importance of 
fully resolving the velocity structures within the lines.  For this reason, we 
constructed our observing program to record spectra of the SMC stars using the 
medium resolution ($\lambda/\Delta\lambda=30,000-45,800$) echelle modes 
(E140M and E230M) of the {\it Space Telescope Imaging Spectrograph\/} (STIS) on 
the HST.  Typical signal-to-noise ratios per resolution element ranged from about 10 
at 1800\,\AA, to 30 at 1300\,\AA, and 40 at 2300\,\AA.

\newpage
\begin{deluxetable*}{
c	
c	
c	
c	
c	
c	
c	
c	
c	
}
\tablewidth{0pt}
\tablecaption{SMC Target Stars\tablenotemark{a}\label{tbl:target_stars}}
\tablehead{
\colhead{Star} &\colhead{AzV\tablenotemark{b}} & \colhead{R.A. (J2000)} & 
\colhead{Dec. (J2000)} &
\colhead{$V$} & \colhead{$(B-V)$} & \colhead{$E(B-V)$} & \colhead{Spectral} &
\colhead{HST Obs.}\\
\colhead{} & \colhead{} & \colhead{($^{\rm h~m~s}$)} & 
\colhead{(\arcdeg~\arcmin~\arcsec)} & \colhead{(mag)} & \colhead{} &
\colhead{Tot/SMC} & \colhead{Type} & \colhead{Pgm(s).\tablenotemark{c}}\\
\colhead{(1)} &
\colhead{(2)} &
\colhead{(3)} &
\colhead{(4)} &
\colhead{(5)} &
\colhead{(6)} &
\colhead{(7)} &
\colhead{(8)} &
\colhead{(9)}
}
\startdata
Sk\,13\dotfill&18 &0 47 12.2& $-73$  \phn6 33&12.44&\phs0.03&0.20/0.16&B2 
Ia&1, 2, 3\\
Sk\,18\dotfill&26&0 47 50.0&$-73$ \phn8 21&12.46&$-0.17$&0.15/0.11&O7 
III&1, 4\\
\nodata&47&0 48 51.5&$-73$ 25 59&13.44&$-0.18$&0.13/0.09&O8 III&1, 5\\
HD\,5045 (Sk\,40)&78&0 50 38.4&$-73$ 28 18&11.05&$-0.05$& 0.14/0.10&B1 
Ia+&1\\
\nodata&80&0 50 43.8&$-72$ 47 42&13.32&$-0.13$& 0.19/0.15&O4-6n(f)p&1, 5\\
\nodata&95&0 51 21.6&$-72$ 44 15&13.78&$-0.18$& 0.14/0.10&O7 III&1, 5\\
\nodata&104&0 51 38.4&$-72$ 48  \phn6&13.17&$-0.16$& 0.06/0.02&B0.5 
Ia&3\\
\nodata&207&0 58 33.2&$-71$ 55 47&14.25&$-0.20$& 0.12/0.09&O7.5 V((f))&1\\
\nodata&216&0 58 59.1&$-72$ 44 34&14.22&$-0.13$& 0.13/0.09&B1 III&3\\
HD\,5980 (Sk\,78)&229&0 59 26.6&$-72$  \phn9 54&11.85&$-0.23$& 0.07/0.04& 
WN6h&6, 7, 8, 9\tablenotemark{d}\\
Sk\,85\dotfill&242&1  \phn0  \phn6.9&$-72$ 13 57&12.07&$-0.12$& 
0.07/0.04&B1 Ia&1\\
\nodata&321&1  \phn2 57.1&$-72$  \phn8  \phn9&13.76&$-0.19$& 0.12/0.09&O9 
IInp&1\\
Sk\,108\dotfill&332&1  \phn3 25.2&$-72$  \phn6 44&12.40&$-
0.24$&\nodata&WN3+O6.5(n)&10, 11\tablenotemark{e}\\
\nodata&388&1  \phn5 39.5&$-72$ 29 27&14.09&$-0.21$& 0.11/0.08&O4 V&1\\
Sk\,143\dotfill&456&1 10 55.8&$-72$ 42 56&12.83&\phs0.10&0.36/0.33&O9.7 
Ib&1, 2\\
\nodata&476&1 13 42.5&$-73$ 17 30&13.52&$-0.09$& 0.23/0.20&O6.5 V&1\\
Sk\,190\dotfill&\nodata&1 31 28.0&$-73$ 22 14&13.54&$-0.18$& 0.11/0.07&O8 
Iaf&1\\
Sk\,191\dotfill&\nodata&1 41 42.1&$-73$ 50 38&11.85&$-0.04$& 0.14/0.10&B1.5 
Ia&1, 3\\

\enddata
\tablenotetext{a}{Data and their sources are taken from Table~1 of Welty et al. 
(2012).}
\tablenotetext{b}{Number in the catalog of Azzopardi \& Vigneau (1982).}
\tablenotetext{c}{Observing program numbers and principal investigators: (1) 
13778 (E.~Jenkins), (2) 9383 (K.~Gordon), (3) 9116 (D.~Lennon), (4) 12978 
(D.~Welty), (5) 7437 (D.~Lennon), (6) 7480 (G.~Koenigsberger), (7) 9094 
(G.~Koenigsberger), (8) 11623 (G.~Koenigsberger), (9) 13373 (G.~Koenigsberger), 
(10) 4048 (J.~Hutchings), (11) 5608 (D.~Welty).}
\tablenotetext{d}{While a detailed analysis of interstellar lines toward this star was 
undertaken by Koenigsberger et al. (2001), their column densities for the SMC 
features were derived using the formula for optically thin absorption, which we feel 
is inappropriate.  Therefore, we derived all of the column densities independently.}
\tablenotetext{e}{Column densities were not derived here, but instead were taken 
from Welty et al. (1997), Mallouris (2003) and Sofia et al. (2006).  When 
appropriate, we adjusted column densities to reflect changes in some $f$-values.}
\end{deluxetable*}

With the exception of one star, AzV\,95, the SMC absorption features were well 
separated from those of the disk and halo of our Galaxy.  Our measurements of the 
lines included heliocentric velocities longward of $+60\,{\rm km~s}^{-1}$, which is 
consistent with the lower velocity limit for the Ti~II lines measured by Welty \& 
Crowther (2010).   The upper velocity limits for most of the absorptions were at 
about  $+190\,{\rm km~s}^{-1}$, but the first 8 stars listed in 
Table~\ref{tbl:target_stars} showed additional separate, narrow absorption at 
higher velocities, most of which reached up to about $+240\,{\rm km~s}^{-1}$.  In 
all cases, we determined column densities over the entire range of SMC velocities, so 
that they could be compared to the total amounts of hydrogen atoms and molecules.

\newpage
\vspace*{-1cm}
\begin{deluxetable*}{
l	
c	
c	
c	
c	
c	
c	
l	
}
\tablewidth{0pt}
\tablecaption{Transitions and SMC Stellar Abundances\label{tbl:trans}}
\tablecolumns{8}
\tablehead{
\colhead{Element} & \colhead{$\lambda$} & \multicolumn{2}{c}{$\log 
f\lambda$\tablenotemark{a}} & 
\colhead{~~~} & \multicolumn{3}{c}{SMC Total Abundances} \\
\cline{3-4} \cline{6-8}
\colhead{$X$} & \colhead{(\AA)} & \colhead{Value} & \colhead{uncertainty} & 
\colhead{} & \colhead{Adopted Reference} & \colhead{Measured Deviations} & 
\colhead{Sources\tablenotemark{c}}\\
&&&&&\colhead{$\log (X/{\rm H})_\sun$\tablenotemark{b}~+~11.35} & 
\colhead{from [$X$/H]\tablenotemark{b}~$-~0.65$}\\
\colhead{(1)} &
\colhead{(2)} &
\colhead{(3)} &
\colhead{(4)} &
\colhead{} &
\colhead{(5)} &
\colhead{(6)} &
\colhead{(7)}
}
\startdata
Mg~II&1240&$-0.355$&0.05\tablenotemark{d}&&6.95&$-
0.21\tablenotemark{e}$&1, 2, 3, 4, 5, 6\\
Si~II&1808&\phs0.575&0.04&&6.86&$-0.09$&1, 2, 3, 4, 5, 6\\
S~II&1251&\phs0.809\tablenotemark{f}&0.04&&6.47&\phs0.03&2,  7,  8, 9\\
&1254&\phs1.113\tablenotemark{f}&0.04&&\\
Ti~II\tablenotemark{g}&\nodata&\nodata&\nodata&&4.30&\phs0.08&8, 10, 11\\
Cr~II&2056&\phs2.326&0.024&&4.99&\phs0.12&7, 8, 10, 11\\
&2066&\phs2.024&0.025&&\\
Mn~II&2577&\phs2.966\tablenotemark{h}&0.006&&4.78&\phs0.23
\tablenotemark{e}&8, 11\\
&2594&\phs2.859\tablenotemark{h}&0.02&&\\
&2606&\phs2.707\tablenotemark{h}&0.021&&\\
Fe~II&1608&\phs1.968&0.02\tablenotemark{d}&&6.85&$-0.03$&2, 4, 10, 12\\
&2374&\phs1.871&0.02&&\\
&2261&\phs0.742&0.03&&\\
&2250&\phs0.612&0.03&&\\
&1611&\phs0.347&0.08&&\\
Ni~II&1370&\phs1.906\tablenotemark{i}&0.04&&5.57&\phs0.12&7, 8, 11\\
&1317&\phs1.876\tablenotemark{i}&0.04&&\\
&1741&\phs1.871&0.04&&\\
&1709&\phs1.743&0.04&&\\
&1752&\phs1.686&0.04&&\\
&1455&\phs1.672&0.12&&\\
Zn~II&2063&\phs2.804\tablenotemark{j}&0.04\tablenotemark{d}&&3.91&
\phs0.08&7, 8\\

\enddata
\tablenotetext{a}{Unless otherwise noted, all values are taken from Morton (2003).}
\tablenotetext{b}{Solar photospheric abundances adopted from values listed by 
Asplund et al. (2009).  The value 11.35 arises from adding our adopted SMC 
logarithmic metallicity $-0.65$ to the standard representations based on H = 12.}
\tablenotetext{c}{(1) Trundle et al. (2004), (2) Hunter et al. (2005), (3) Dufton et al. 
(2005), (4) Trundle et al. (2007), (5) Hunter et al. (2007), (6) Hunter et al. (2009), 
(7) Luck \& Lambert (1992), (8) Luck et al. (1998), (9) Rolleston et al. (2003), (10) 
Venn (1999), (11) Russell \& Dopita (1992), (12) Korn et al. (2000).}
\tablenotetext{d}{This uncertainty is our own estimate, since no value was specified 
by Morton (2003).}
\tablenotetext{e}{See the text in Section~\protect\ref{sec:stellar_abundances}.}
\tablenotetext{f}{These lines are sometimes badly saturated.  For cases where the 
saturation is not severe, they were analyzed using the technique devised by Jenkins 
(1996) to correct for unresolved saturated components.  Revised $f$-values are 
from Kisielius et al. (2014).}
\tablenotetext{g}{Column densities are taken from Welty \& Crowther (2010), with 
uncertainties scaled according to the relative errors in the equivalent width 
measurements.}
\tablenotetext{h}{Revised $f$-values are from Den Hartog et al. (2011).}
\tablenotetext{i}{Revised $f$-values are from Jenkins \& Tripp (2006).}
\tablenotetext{j}{Revised $f$-values are from Kisielius et al. (2015).  We corrected 
for interference from an almost coincident Cr~II line (2062.234\,\AA, displaced 
from the Zn~II line by $\Delta v=-62\,\kms$) using the apparent optical depth 
information from another Cr~II line at 2056\,\AA.}
\end{deluxetable*}
\newpage

Table~\ref{tbl:trans} lists the transitions that were suitable for deriving column 
densities of various elements in their preferred stages of ionization for H~I regions.  
The oscillator strengths ($f$-values) for most transitions were taken from the 
compilation of Morton (2003), but with a few exceptions noted in the footnotes of 
the table.  We derived column densities from the apparent optical depths (AOD) 
(Savage \& Sembach 1991) after the local parts of the spectra were renormalized to 
continuum levels defined by best-fitting Legendre polynomials (Sembach \& Savage 
1992).   In almost all cases, our analysis for transitions of different strengths for a 
single element yielded outcomes that agreed with each other to within the 
measurement uncertainties.  Nevertheless, we acknowledge that our ability to 
uncover evidence of hidden saturated absorptions was somewhat hampered by our 
not being able to simultaneously measure transitions whose strengths differed by 
more than a factor of two.

\begin{figure}[b]
\plotone{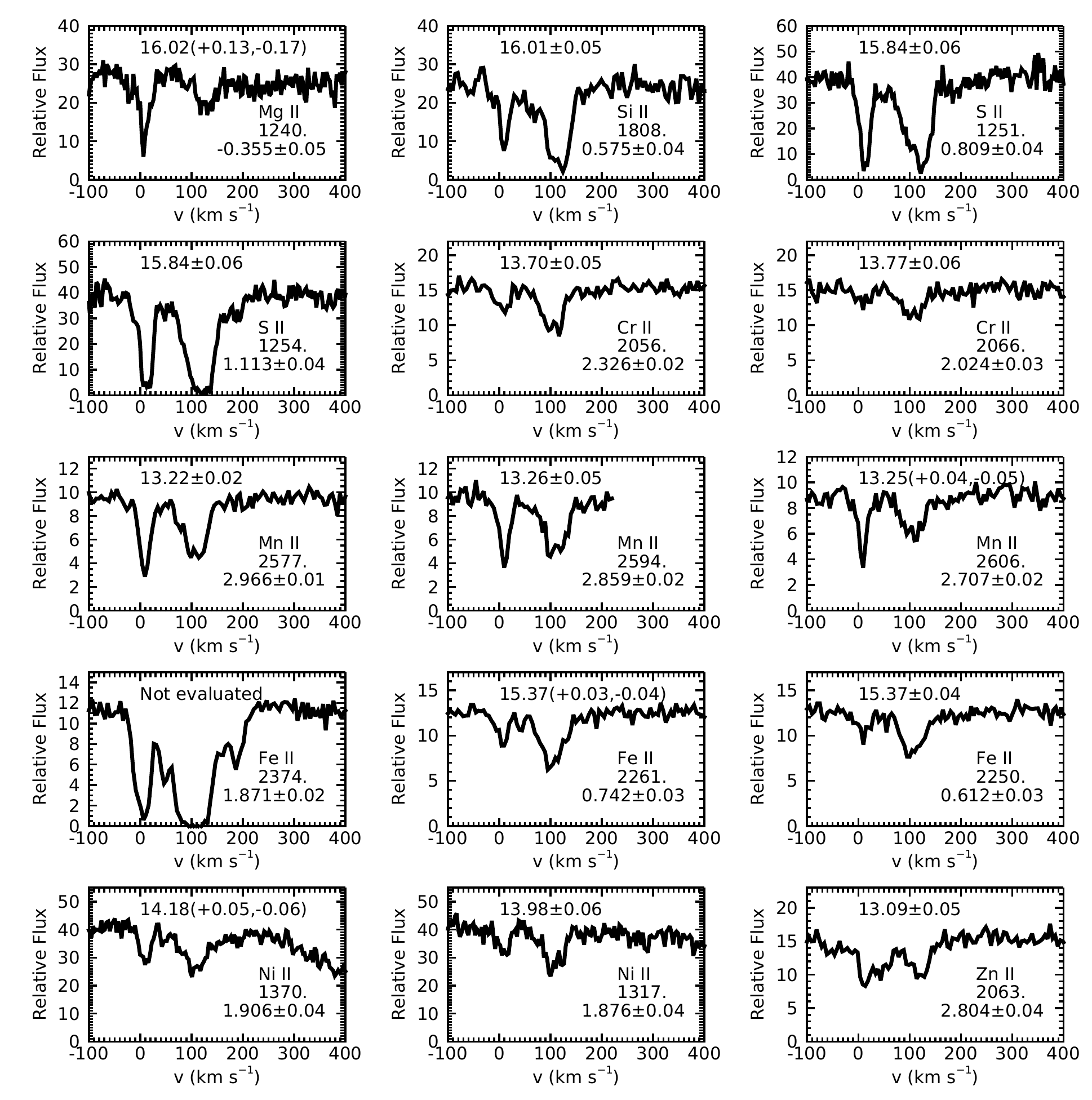}
\caption{Absorption features for various elements in the spectrum of AzV\,95, with 
the recorded intensities plotted as a function of heliocentric velocities.  Near the top 
of each panel, we show our outcome for $\log (N)$ and its uncertainty limits.  In the 
lower portion of each panel, we list the ion, together with the wavelength and value 
for $\log f\lambda$ for the transition.  For this star, the outcomes for Ni represent 
the only case where the results for a single element differed by more than the 
uncertainties that we computed.  The outcomes for the two transitions of S~II are 
identical because both lines were analyzed together using the correction method of 
Jenkins (1996) for interpreting apparent optical depths that have unresolved 
saturated structures.  We show the absorption by the very strong transition of Fe~II 
at 2374\,\AA\ simply to illustrate the range of velocities of even the smallest 
amounts of SMC gas for this star, some of which exhibits an overlap in velocity with 
material in our Galaxy.  (This situation occurred only for this star however.)  The 
absorption at 2374\,\AA\  was much too saturated to derive a column 
density.\label{fig:AzV95_spectra}}
\end{figure}
Figure~\ref{fig:AzV95_spectra} shows the available absorption features for one of 
the stars in our survey, AzV\,95, which is a typical case from the standpoint of the 
star’s brightness, hydrogen column density and our spectral coverage.  Many 
transitions of other elements within our wavelength coverage could not be used.  
For instance, the strong transitions of C~II, N~I, O~I and Si~II were always 
completely saturated.  Even the weak line of Si~II at 1808\,\AA\ was too saturated 
for deriving column densities for some stars.  The two weaker lines (1251 and 
1254\,\AA) of the S~II triplet were sometimes only moderately saturated, which 
made them good candidates for analyzing their apparent optical depths after 
correcting for unresolved saturated structures using the method proposed by 
Jenkins (1996).  At the other extreme, the intersystem lines of C~II (2325\,\AA), 
Si~II (2335\,\AA), and usually O~I (1356\,\AA) were too weak to observe.  
Likewise, with very few exceptions we found that the allowed transitions of B~II, 
Cu~II, Ge~II Ga~II, and Kr~I were too weak to measure above the noise.  A feature 
of P~II at 1152.82\,\AA\ had a respectable strength, but for most stars our spectra 
at that wavelength had a very poor signal-to-noise ratio.  Thus the outcomes would 
be uncertain and could also be subject to systematic errors (Fox et al. 2005).

For some transitions, we had to acknowledge and compensate for overlaps in 
absorption: the Mg~II line at 1239.925\,\AA\ covering the velocity range of the 
SMC gas suffered interference from the Milky Way absorption arising the other 
member of the doublet at 1240.395\,\AA.  Thus, for deriving the column density of 
Mg~II in the SMC, we had to rely entirely on the line at the longer wavelength.  The 
transition of Zn~II at 2026.137\,\AA\ suffers interference from nearby lines of 
Cr~II (2026.269\,\AA), Co~II (2026.412\,\AA) and Mg~I (2026.477\,\AA).  The 
Cr~II and Co~II lines are probably too weak to matter, but the Mg~I line may make 
a non-negligible contribution.  Thus, we chose not to use the 2026\,\AA\ line of 
Zn~II, chiefly because we could not correct for the Mg~I absorption (another 
transition of Mg~I at 2852.963\,\AA\ that could have been used as a reference was 
outside our wavelength coverage).  The other available line of Zn~II at 
2062.660\,\AA\ also has interference from a Cr~II transition (2062.234\,\AA).  
Under most circumstances, the wavelength separation of the two lines is large 
enough to not cause a problem.  However the range of velocities over which SMC 
absorptions occur result in some overlap.   Hence, we had to subtract off the optical 
depths of the Cr~II feature from those of Zn~II, using as a reference rescaled 
versions of the optical depths that we obtained from the Cr~II absorption at 
2056.257\,\AA.

For the column density outcome arising from any single transition, we assigned an 
error $\sigma$ that included uncertainties caused by (1) the noise in the absorption 
profile, (2) errors in defining the continuum level, and (3) the uncertainty of the
$f$-value (see Column~4 of Table~\ref{tbl:trans}), all of which were combined in 
quadrature.  Continuum uncertainties can have a large impact on the measurement 
accuracies of weak lines.   We evaluated them from the expected formal 
uncertainties in the polynomial coefficients of the fit, as described by Sembach \& 
Savage (1992), and then we multiplied them by 2 in order to make an approximate 
allowance for additional uncertainties arising from our freedom in selecting the 
most appropriate polynomial order.   We determined the effects of such 
uncertainties by evaluating the column densities at the lower and upper extremes 
for the continuum.  When results from more than one transition of a given element 
were available, we evaluated a column density based on a weighted average, with 
the weights proportional to the respective inverse square errors $1/\sigma_i^2$.  
The uncertainty in any such combined result was assigned a value equal to 
$\left(\sum_i \sigma_i^{-2}\right)^{-0.5}$.  For the measurements of S~II, we 
added in quadrature an additional error of 0.1~dex to account for possible 
inaccuracies in correcting for saturation.\footnote{Except for AzV\,332, where we 
imported the value for $N$(S~II) from Mallouris (2003).}

\section{STELLAR REFERENCE ABUNDANCES}\label{sec:stellar_abundances}

As we had discussed in Section~\ref{sec:motivation}, a key component of our 
investigation is the establishment of a set of reference element abundances against 
which we can compare our interstellar gas-phase abundances.  The findings for 
stars that have recently formed out of the ISM of the SMC should serve as our best 
examples.  We have avoided using abundances derived from emission lines in H~II 
regions because dust grains could survive within them and hence their gas-phase 
abundances could be reduced.

We have examined about 30 papers that presented results for stellar abundances in 
the SMC.\footnote{In the appendix of the paper by Welty et al. (1997), there is a 
good summary of prior abundance determinations in the SMC.}   In narrowing the 
selection and making choices for defining our adopted abundances (which varied 
from one element to the next), we have favored results that came from relatively 
recent investigations that have had access to modern, large-aperture telescopes 
with more powerful spectrographs, improved stellar atmosphere and line-formation 
codes (we preferred NLTE calculations over LTE ones), and more accurate 
transition strengths.   We have also recognized that the spectra of some types of 
stars are easier to interpret than others.  For instance, main-sequence and giant
B-type stars are relatively straightforward to analyze, while Cepheids have the 
complications of differential velocities in the pulsations that make the lines hard to 
interpret, and cool stars with surface convection require more elaborate modeling.   

Some elements appear to have a small dispersion in the outcomes from different 
investigations ($\sigma \sim 0.15$), such as Mg, Si, Ti, and Cr, while the elements 
Mn, Fe, Ni and Zn either have few determinations or larger dispersions in the 
outcomes.  Low dispersions of results do not necessarily indicate that the average 
results are reliable, since systematic errors could persist throughout many or all of 
the investigations (Blanco-Cuaresma et al. 2016 ; Hinkel et al. 2016).   For instance, 
most of the stellar abundances in the SMC must be derived from spectra of luminous 
stars, for which the uncertainties of the model stellar atmospheres dominate the 
total uncertainties of the abundance determinations.   

The elements in Table~\ref{tbl:trans} can be divided into the $\alpha$ group; Mg, 
Si, S, and Ti and the Fe-peak species from Cr to Zn. The dominant isotopes of the 
$\alpha$ elements have nuclei with an integral number of $\alpha$ particles, except 
for Ti, which also has a clump of 4 neutrons. In old, metal-poor stars, such as those 
in globular clusters and the Galactic thick disc and halo, there is a gradual increase 
of $\alpha$ elements relative to Fe as [Fe/H] diminishes from 0.0 to about $-1.0$.  
However, Mucciarelli (2014) found that this trend is displaced toward lower 
metallicities in the SMC, and that for $[{\rm Fe/H}]\approx -0.7$, $[\alpha/{\rm 
Fe}]\approx 0$.  If we take a straight mean of the 4 $\alpha$ elements in 
Table~\ref{tbl:trans} we find that [$\alpha$/H]~=~$-0.7$.\footnote{While O is not 
listed in Table~\protect\ref{tbl:trans}, an average value for [O/H]~=~$-0.65$ is 
nearly consistent with the abundance deficiencies of other $\alpha$ elements.}   
This average does not include C, which appears to have a much lower abundance 
(this is an issue that we will discuss in Section~\ref{sec:carbon}).  For the 5 Fe-peak 
elements a straight mean of [Fe/H]~=~$-0.60$. Of the 5 Fe-peak elements the 
abundances of Mn and Zn are the least certain.  Mn has relatively few 
determinations. The abundance of Zn, which is featured in many studies of DLA gas 
compositions, depends on 2 lines of Zn~I in Cepheids, the atmospheres of which are 
difficult to model because of the running wave that passes through them during 
each cycle. Even so, omitting Mn and Zn from the mean for the Fe group leaves it 
virtually unchanged at $-0.58$.

%
%
\floattable
\begin{deluxetable}{
l 	
c 	
c 	
c 	
c 	
c 	
c 	
c 	
c 	
c 	
}
\rotate
\tablewidth{0pt}
\tablecaption{SMC Column Densities and Depletions\label{tbl:col_dens_depl1}}
\tablehead{
\colhead{Element} & 
\colhead{AzV\,18} & \colhead{AzV\,26} & \colhead{AzV\,47} & 
\colhead{AzV\,78} & \colhead{AzV\,80} & \colhead{AzV\,95} & 
\colhead{AzV\,104} & \colhead{AzV\,207} & \colhead{AzV\,216}\\
\colhead{(1)} &
\colhead{(2)} &
\colhead{(3)} &
\colhead{(4)} &
\colhead{(5)} &
\colhead{(6)} &
\colhead{(7)} &
\colhead{(8)} &
\colhead{(9)} &
\colhead{(10)}
}
\startdata

$\log N({\rm H~I})$&
$22.04\pm0.02$& 
$21.70^{+0.05}_{-0.06}$& 
$21.32\pm0.04$& 
$21.70^{+0.05}_{-0.06}$\tablenotemark{a}& 
$21.81\pm0.02$& 
$21.49\pm0.04$& 
$21.45^{+0.05}_{-0.06}$& 
$21.43^{+0.05}_{-0.06}$& 
$21.64\pm0.03$ 

\\

$\log N({\rm H_2})$&
$20.36^{+0.07}_{-0.08}$& 
$20.63^{+0.05}_{-0.06}$& 
$18.54^{+0.4}_{-1.0}$& 
$18.64^{+0.2}_{-0.4}$\tablenotemark{b}& 
$20.08^{+0.2}_{-0.4}$& 
$19.40^{+0.08}_{-0.10}$& 
$19.23^{+0.2}_{-0.4}$& 
$19.40^{+0.09}_{-0.11}$& 
$18.78^{+0.3}_{-2.3}$ 

\\

$\log N({\rm H_{tot}})$&
$22.06\pm0.02$& 
$21.77\pm0.04$&
$21.32\pm0.04$&
$21.70^{+0.05}_{-0.06}$& 
$21.83\pm0.02$&
$21.50\pm0.04$&
$21.46^{+0.05}_{-0.06}$& 
$21.44^{+0.05}_{-0.06}$& 
$21.64\pm0.03$

\\ [5pt]

$\log N({\rm Mg~II})$&
$16.68^{+0.06}_{-0.07}$& 
$16.11^{+0.04}_{-0.05}$&  
$16.07^{+0.10}_{-0.13}$&  
$16.26^{+0.09}_{-0.12}$&  
$16.37^{+0.08}_{-0.10}$&  
$16.02^{+0.12}_{-0.17}$&  
$16.14^{+0.09}_{-0.11}$&  
\nodata& 
 $16.33^{+0.07}_{-0.09}$ 

\\
$D$(Mg)
&$-0.33^{+0.08}_{-0.10}$ 
&$-0.61^{+0.08}_{-0.09}$ 
&$-0.21^{+0.11}_{-0.13}$ 
&\nodata 
&$-0.40^{+0.09}_{-0.11}$ 
&$-0.43^{+0.13}_{-0.18}$ 
&$-0.26^{+0.10}_{-0.12}$ 
&\nodata 
&$-0.26^{+0.08}_{-0.09}$ 
\\ [5pt]

$\log N({\rm Si~II})$&
\nodata& 
$16.12^{+0.04}_{-0.05}$&  
$15.88^{+0.05}_{-0.06}$&  
 \nodata&
\nodata& 
$16.01^{+0.05}_{-0.06}$&  
$15.96^{+0.09}_{-0.12}$&  
$16.02^{+0.16}_{-0.27}$&  
\nodata 

 \\
$D$(Si)
&\nodata 
&$-0.51\pm 0.06$ 
&$-0.30\pm0.07$ 
&\nodata 
&\nodata 
&$-0.35\pm 0.07$ 
&$-0.35^{+0.11}_{-0.13}$ 
&$-0.28^{+0.17}_{-0.27}$ 
&\nodata 
\\ [5pt] 

$\log N({\rm S~II})$& 
\nodata& 
\nodata& 
$15.84^{+0.07}_{-0.08}$&  
\nodata& 
\nodata& 
$15.84^{+0.06}_{-0.07}$&  
$15.78\pm 0.05$&  
\nodata& 
\nodata 

\\
$D$(S) 
&\nodata 
&\nodata 
&$ 0.05^{+0.13}_{-0.14}$ 
&\nodata 
&\nodata 
&$-0.13^{+0.12}_{-0.13}$ 
&$-0.15\pm 0.12$ 
&\nodata 
&\nodata 
\\ [5pt] 

$\log N({\rm Ti~II})$&
$12.98\pm0.01$&  
$12.90\pm0.01$& 
$12.82\pm 0.03$& 
$12.92\pm0.01$& 
$12.81\pm 0.03$& 
$12.82\pm 0.04$& 
\nodata& 
$12.49^{+0.07}_{-0.08}$& 
\nodata 

 \\
$D$(Ti)
&$-1.38\pm 0.02$ 
&$-1.17^{+0.05}_{-0.04}$ 
&$-0.80\pm 0.05$ 
&$-1.08^{+0.06}_{-0.05}$ 
&$-1.32^{+0.03}_{-0.04}$ 
&$-0.98\pm 0.06$ 
&\nodata 
&$-1.25\pm 0.09$ 
&\nodata 
\\ [5pt]

$\log N({\rm Cr~II})$& 
$13.97^{+0.05}_{-0.06}$& 
$13.84\pm 0.02$& 
$13.69\pm 0.04$& 
$14.12^{+0.04}_{-0.05}$& 
$13.72\pm0.03$& 
$13.72\pm 0.04$& 
$13.50^{+0.08}_{-0.09}$& 
$13.52^{+0.05}_{-0.06}$& 
$13.69^{+0.04}_{-0.05}$ 

\\

$D$(Cr)
&$-1.07\pm 0.06$ 
&$-0.92\pm 0.05$ 
&$-0.62^{+0.06}_{-0.05}$ 
&$-0.57\pm 0.07$ 
&$-1.10\pm 0.04$ 
&$-0.76\pm 0.06$ 
&$-0.95^{+0.09}_{-0.10}$ 
&$-0.91\pm 0.07$ 
&$-0.94\pm 0.05$ 
\\ [5pt] 

$\log N({\rm Mn~II})$& 
$13.47\pm 0.02$& 
$13.37\pm 0.01$& 
$13.24\pm 0.03$& 
$13.37\pm 0.01$& 
$13.25\pm 0.02$& 
$13.23\pm 0.02$& 
\nodata& 
$13.08\pm 0.03$& 
\nodata 

\\

$D$(Mn) 

&$-1.37\pm 0.03$ 
&$-1.17\pm 0.05$ 
&$-0.86\pm 0.05$ 
&$-1.12^{+0.06}_{-0.05}$ 
&$-1.35\pm 0.03$ 
&$-1.05^{+0.05}_{-0.04}$ 
&\nodata 
&$-1.13\pm 0.06$ 
&\nodata 
\\ [5pt] 

$\log N({\rm Fe~II})$& 
$15.61\pm 0.03$& 
$15.51\pm 0.02$& 
$15.37\pm 0.03$& 
$15.48\pm 0.03$& 
$15.39\pm 0.03$& 
$15.37\pm 0.03$& 
$15.31^{+0.03}_{-0.04}$& 
$15.16\pm 0.04$& 
$15.41\pm 0.03$ 

\\

$D$(Fe) 
&$-1.29^{+0.03}_{-0.04}$ 
&$-1.11\pm 0.05$ 
&$-0.81\pm 0.05$ 
&$-1.07\pm 0.06$ 
&$-1.29\pm 0.03$ 
&$-0.98\pm 0.05$ 
&$-0.99\pm 0.06$ 
&$-1.13^{+0.07}_{-0.06}$ 
&$-1.08\pm 0.04$ 
\\ [5pt]

$\log N({\rm Ni~II})$& 
$14.31^{+0.03}_{-0.04}$& 
$14.25\pm 0.02$& 
$14.08\pm 0.04$& 
$14.19\pm 0.03$& 
$14.11\pm 0.04$& 
$14.06\pm 0.04$& 
$14.03\pm 0.04$& 
\nodata& 
$14.16^{+0.031}_{-0.044}$ 

\\

$D$(Ni) 
&$-1.32\pm 0.04$ 
&$-1.09\pm 0.05$ 
&$-0.81\pm 0.06$ 
&$-1.08\pm 0.06$ 
&$-1.29^{+0.04}_{-0.05}$ 
&$-1.01\pm 0.06$ 
&$-1.00^{+0.07}_{-0.06}$ 
&\nodata 
&$-1.05\pm 0.04$ 
\\ [5pt] 

$\log N({\rm Zn~II})$& 
$13.67^{+0.07}_{-0.08}$& 
$13.26^{+0.04}_{-0.05}$& 
$13.09\pm 0.06$& 
\nodata& 
$13.23\pm 0.05$& 
$13.09^{+0.05}_{-0.06}$& 
$12.97^{+0.13}_{-0.18}$& 
$13.14\pm 0.05$& 
$13.28^{+0.06}_{-0.07}$ 

\\

$D$(Zn)
&$-0.29^{+0.07}_{-0.09}$ 
&$-0.42\pm 0.06$ 
&$-0.14^{+0.07}_{-0.08}$ 
&\nodata 
&$-0.51\pm 0.05$ 
&$-0.32\pm 0.07$ 
&$-0.40^{+0.14}_{-0.19}$ 
&$-0.21\pm 0.07$ 
&$-0.27\pm 0.07$ 

\\

\enddata
\tablenotetext{a}{Value determined from our own spectrum by finding the 
optimum value for $e^{+\tau({\rm L}\alpha)}$ that cancels the interstellar 
absorption; see, e.g., Bohlin (1975).  The value published by Welty et al. (2012) has a 
large uncertainty, since it was derived from IUE data.}
\tablenotetext{b}{No observation of H$_2$ available for this star.  The value shown 
here is for AzV\,81 which is 1\farcm2 away from AzV\,80.}
\end{deluxetable}
\addtocounter{table}{-1}
%
%
\floattable
\begin{deluxetable}{
l 	
c 	
c 	
c 	
c 	
c 	
c 	
c 	
c 	
c 	
}
\rotate
\tablecaption{SMC Column Densities and Depletions 
(continued)\label{tbl:col_dens_depl2}}
\tablehead{
\colhead{Element} &
\colhead{AzV\,229} & \colhead{AzV\,242} & \colhead{AzV\,321} & 
\colhead{AzV\,332} & \colhead{AzV\,388} & \colhead{AzV\,456} & 
\colhead{AzV\,476} & \colhead{Sk\,190} & \colhead{Sk\,191}\\ 
\colhead{(1)} &
\colhead{(2)} &
\colhead{(3)} &
\colhead{(4)} &
\colhead{(5)} &
\colhead{(6)} &
\colhead{(7)} &
\colhead{(8)} &
\colhead{(9)} &
\colhead{(10)}
}
\startdata


$\log N({\rm H~I})$&
$21.06\pm0.04$& 
$21.32\pm0.04$& 
$20.70^{+0.10}_{-0.05}$\tablenotemark{a}& 
$20.54^{+0.13}_{-0.19}$& 
$21.18\pm0.04$& 
$21.00^{+0.05}_{-0.06}$& 
$21.85^{+0.06}_{-0.07}$& 
$20.62\pm0.04$& 
$21.51\pm0.04$

\\

$\log N({\rm H_2})$&
$15.66^{+0.2}_{-0.4}$&
$17.21^{+0.3}_{-2.3}$&
$14.44^{+0.3}_{-2.3}$&
$14.50^{+0.10}_{-0.13}$&
$19.40^{+0.09}_{-0.11}$&
$20.93^{+0.09}_{-0.11}$&
$20.95^{+0.20}_{-0.38}$&
$17.42^{+0.06}_{-0.07}$&
$20.65^{+0.10}_{-0.13}$

\\
$\log N({\rm H_{tot}})$&
$21.06\pm0.04$&
$21.32\pm0.04$&
$20.70^{+0.10}_{-0.05}$&
$20.54^{+0.13}_{-0.19}$&
$21.19\pm0.04$&
$21.43^{+0.06}_{-0.07}$&
$21.95^{+0.07}_{-0.08}$&
$20.62\pm0.04$&
$21.62\pm0.04$

\\ [5pt]

$\log N({\rm Mg~II})$&
$15.82^{+0.09}_{-0.11}$&  
$16.03^{+0.09}_{-0.11}$& 
\nodata& 
 \nodata& 
$16.08^{+0.11}_{-0.15}$&  
\nodata& 
\nodata& 
 \nodata&
 \nodata 

\\

$D$(Mg)
&$-0.20^{+0.10}_{-0.12}$ 
&$-0.24^{+0.10}_{-0.11}$ 
&\nodata 
&\nodata 
&$-0.07^{+0.12}_{-0.15}$ 
&\nodata 
&\nodata 
&\nodata 
&\nodata 
 \\ [5pt]

$\log N({\rm Si~II})$&
$15.74^{+0.04}_{-0.05}$&  
$15.89^{+0.05}_{-0.06}$&  
$15.47^{+0.08}_{-0.09}$&  
$15.52^{+0.05}_{-0.06}$&  
$15.68^{+0.05}_{-0.06}$&  
$15.51^{+0.09}_{-0.11}$&  
\nodata& 
$15.59^{+0.08}_{-0.10}$&  
$15.24^{+0.08}_{-0.09}$ 

 \\

$D$(Si)
&$-0.18\pm 0.06$ 
&$-0.29\pm 0.07$ 
&$-0.09^{+0.09}_{-0.13}$ 
&$ 0.12^{+0.19}_{-0.14}$ 
&$-0.38\pm 0.07$ 
&$-0.78^{+0.11}_{-0.12}$ 
&\nodata 
&$ 0.11^{+0.09}_{-0.11}$ 
&$-1.24^{+0.09}_{-0.10}$ 
 \\ [5pt] 

$\log N({\rm S~II})$&
$15.57^{+0.04}_{-0.05}$& 
$15.83\pm 0.05$&  
$15.46\pm 0.05$&  
$15.35\pm 0.06$& 
$15.71\pm 0.05$& 
\nodata& 
\nodata& 
$15.32\pm 0.05$& 
$15.28^{+0.07}_{-0.08}$ 

 \\

$D$(S)
&$ 0.04\pm 0.12$ 
&$ 0.04\pm 0.12$ 
&$ 0.29^{+0.12}_{-0.15}$ 
&$ 0.34^{+0.19}_{-0.14}$ 
&$ 0.05\pm 0.12$ 
&\nodata 
&\nodata 
&$ 0.23\pm 0.12$ 
&$-0.80\pm 0.13$ 
\\ [5pt] 

$\log N({\rm Ti~II})$&
$12.38\pm 0.04$& 
$12.50\pm 0.02$\tablenotemark{b}& 
$12.10\pm 0.03$& 
\nodata& 
$12.14^{+0.07}_{-0.09}$& 
$12.01^{+0.09}_{-0.11}$& 
$12.82\pm 0.01$& 
$12.14^{+0.08}_{-0.09}$& 
$11.62^{+0.08}_{-0.10}$\tablenotemark{c} 

 \\

$D$(Ti)
&$-0.98\pm 0.06$ 
&$-1.12\pm 0.05$ 
&$-0.90^{+0.06}_{-0.10}$ 
&\nodata 
&$-1.35^{+0.08}_{-0.09}$ 
&$-1.72^{+0.11}_{-0.13}$ 
&$-1.43^{+0.08}_{-0.07}$ 
&$-0.78^{+0.09}_{-0.10}$ 
&$-2.29^{+0.09}_{-0.11}$ 
\\ [5pt]

$\log N({\rm Cr~II})$&
$13.33^{+0.07}_{-0.08}$& 
$13.26^{+0.16}_{-0.25}$& 
\nodata& 
$13.13\pm 0.03$& 
\nodata& 
\nodata& 
$13.75\pm 0.03$& 
\nodata& 
\nodata 

 \\

$D$(Cr)
&$-0.72^{+0.08}_{-0.09}$ 
&$-1.05^{+0.16}_{-0.25}$ 
&\nodata 
&$-0.40^{+0.19}_{-0.13}$ 
&\nodata 
&\nodata 
&$-1.19^{+0.08}_{-0.07}$ 
&\nodata 
&\nodata 
\\ [5pt] 

$\log N({\rm Mn~II})$&
$12.86\pm 0.03$& 
$12.95\pm 0.04$& 
$12.65^{+0.10}_{-0.13}$& 
$12.82\pm 0.04$& 
$12.72^{+0.08}_{-0.09}$& 
$12.67^{+0.07}_{-0.08}$& 
$13.33\pm 0.02$& 
\nodata& 
$12.35^{+0.09}_{-0.11}$ 

 \\

$D$(Mn)
&$-0.98\pm 0.05$ 
&$-1.15\pm 0.06$ 
&$-0.83^{+0.11}_{-0.17}$ 
&$-0.50^{+0.19}_{-0.14}$ 
&$-1.25^{+0.09}_{-0.11}$ 
&$-1.55\pm 0.10$ 
&$-1.39^{+0.08}_{-0.07}$ 
&\nodata 
&$-2.05^{+0.10}_{-0.11}$ 
\\ [5pt] 

$\log N({\rm Fe~II})$&
$14.88\pm0.02$& 
$15.08\pm 0.04$& 
$14.66\pm 0.03$& 
$14.82\pm 0.04$& 
$14.74\pm 0.02$& 
$14.60^{+0.04}_{-0.05}$& 
$15.43^{+0.02}_{-0.03}$& 
$14.57^{+0.02}_{-0.03}$& 
$14.28\pm 0.03$ 

\\

$D$(Fe)
&$-1.03^{+0.05}_{-0.04}$ 
&$-1.09\pm 0.06$ 
&$-0.89^{+0.06}_{-0.10}$ 
&$-0.57^{+0.19}_{-0.14}$ 
&$-1.30^{+0.05}_{-0.04}$ 
&$-1.68\pm 0.08$ 
&$-1.37^{+0.08}_{-0.07}$ 
&$-0.90\pm 0.05$ 
&$-2.18\pm 0.05$ 
\\ [5pt]

$\log N({\rm Ni~II})$&
$13.76^{+0.03}_{-0.04}$& 
$13.87^{+0.04}_{-0.05}$& 
$13.59^{+0.11}_{-0.14}$& 
$13.47^{+0.10}_{-0.13}$& 
\nodata& 
\nodata& 
$14.12^{+0.06}_{-0.07}$& 
$13.44^{+0.09}_{-0.12}$& 
\nodata 

 \\

$D$(Ni)
&$-0.87^{+0.06}_{-0.05}$ 
&$-1.03\pm 0.06$ 
&$-0.68^{+0.12}_{-0.17}$ 
&$-0.64^{+0.21}_{-0.18}$ 
&\nodata 
&\nodata 
&$-1.40\pm 0.10$ 
&$-0.75^{+0.10}_{-0.12}$ 
&\nodata 
\\ [5pt]
 
$\log N({\rm Zn~II})$&
$12.65^{+0.07}_{-0.08}$& 
$13.39^{+0.14}_{-0.21}$& 
$12.71^{+0.10}_{-0.12}$& 
$12.53\pm 0.05$& 
$12.88^{+0.06}_{-0.08}$& 
$13.11^{+0.07}_{-0.08}$& 
$13.39^{+0.13}_{-0.20}$& 
\nodata& 
\nodata 

 \\

$D$(Zn)
&$-0.32^{+0.08}_{-0.09}$ 
&$ 0.16^{+0.15}_{-0.21}$ 
&$ 0.10^{+0.11}_{-0.16}$ 
&$ 0.08\pm 0.23$ 
&$-0.23^{+0.08}_{-0.09}$ 
&$-0.23\pm 0.10$ 
&$-0.46^{+0.09}_{-0.08}$ 
&\nodata 
&\nodata 

\\

\enddata
\tablenotetext{a}{Value determined from our own spectrum by finding the 
optimum value for $e^{+\tau({\rm L}\alpha)}$ that cancels the interstellar 
absorption; see, e.g., Bohlin (1975).  The value published by Welty et al. (2012) is 
not as accurate, since it was determined using L$\beta$ absorption.}
\tablenotetext{b}{From D.~Welty (private communication)}
\tablenotetext{c}{Our own evaluation from a figure shown in Cox et al. (2007), but 
with a higher zero level for $N_a(v)$.}
\end{deluxetable}

In short, we feel that it is difficult to obtain elemental abundances in stars with an 
indisputable accuracy better than 0.1~dex.  For this reason, we can simplify our 
reference standards by adopting for all elements a mean value of $-0.65$ for 
[$X$/H] that we obtained for the averages of the $\alpha$ and Fe-peak 
groups.\footnote{The source of Zn in the Galaxy is uncertain because it may be 
produced in either SN\,Ia, SN\,II, or by neutron capture in massive stars (Woosley et 
al. 2002 ; Sukhbold et al. 2016).  Elements heavier than Zn are formed by neutron 
capture so their abundances may differ from those of the iron-group and the alpha 
elements.  Unfortunately, their abundances are too low for us to measure in the 
spectra of our SMC stars.}  These reference values are listed in Column~5 of 
Table~\ref{tbl:trans}.  Deviations of the stellar measurements from our adopted 
reference abundances are listed in Column~6 of the table, followed by a column that 
indicates the sources of the averages in each case.  The largest deviations are found 
for Mg and Mn.  A deficiency for [Mg/Fe] is seen for nearby dwarf spheroidal 
galaxies for $[{\rm Fe/H}]\gtrsim -1.0$ (Tolstoy et al. 2009), but this element is 
usually analyzed using only a blend of Mg~II lines at 4481\,\AA\ (e.g., Trundle et al. 
2007), so its accuracy may not be as good as that of the other elements.  The few 
sources of information on Mn yielded an average $[{\rm Mn/H}]=-0.42$, which 
seems too high relative to other Fe-group elements.  There are only a few lines of 
Mn~II, and there are uncertainties for Mn~I due to departures from LTE in their 
excitation and ionization.  Also Mn is an odd element and the iron group is 
dominated by even elements.  One expects to have $[{\rm Mn/Fe}]>0$ only for 
extremely metal-poor stars.

While we recognized 4 different evaluations of the sulfur abundance, and the 
dispersion between them is moderate, we nevertheless find that the average result 
yields an outcome that is significantly lower than some of our interstellar 
abundances.  We return to this issue later in this paper (Section~\ref{sec:sulfur}).

Overall, to within the expected accuracy of the SMC stellar abundance 
determinations, we find that the Fe-peak elements appear to have about the same 
deficiencies below the solar abundance as the $\alpha$ elements.  Thus, we can 
regard the SMC gas as having a dilute concentration of heavy elements, but with a 
relative pattern from one element to the next that does not seem to differ 
appreciably from the solar one.
\newpage
\section{DEPLETION OUTCOMES}\label{sec:depl_outcomes}

We define the element depletions for the ISM in the SMC in terms of how far the
gas-phase concentrations relative to hydrogen are deficient when compared to the 
stellar abundances that we discussed in Section~\ref{sec:stellar_abundances}.  We 
express these depletions $D(X)$ for an element $X$ according to the 
relation,\footnote{We depart from the notation for a depletion $[X_{\rm gas}/{\rm 
H}]$ that was adopted by Jenkins (2009) for gas in the Milky Way to avoid a possible 
misunderstanding about the use of solar abundances as a comparison standard.  In 
our case, we are comparing interstellar abundances in the SMC with those of stars in 
the SMC.}
\begin{equation}\label{eqn:depletion}
D(X)=\log N(X)-\log N({\rm H_{tot}})-\log(X/{\rm H})_{\rm stellar}~,
\end{equation}
where $N({\rm H_{tot}})=N({\rm H~I})+2N({\rm H_2})$, and values (+12) for our 
adopted stellar abundances $\log(X/{\rm H})_{\rm stellar}$ are listed in Column~5 
of Table~\ref{tbl:trans}.

The topmost portion of Table~\ref{tbl:col_dens_depl1} shows the logarithms of the 
H~I, H$_2$, and H$_{\rm tot}$ column densities, most of which were taken from the 
survey of Welty et al. (2012).   The H~I column densities were derived from 
observations of the damping wings of the L$\alpha$ (and sometimes L$\beta$) 
absorption features in the spectra of the stars.  The outcomes from these 
measurements include contributions from our Galaxy, which cannot be separated 
from the SMC components.  Hence, an allowance for the foreground Galactic 
contribution had to be made on the basis of 21-cm measurements in the same 
directions (Kalberla et al. 2010), using the assumption that the emission was 
optically thin.  Here, the velocity range of Galactic gas was clearly separable from the 
SMC contributions.   These foreground column densities ranged from 3.3 to 
$7.4\times 10^{20}\,{\rm cm}^{-2}$, which constitutes a sizable fraction of the 
totals for the few SMC sight lines with the lowest values.   Welty et al. (2012) 
determined $N$(H$_2$) from their Voigt profile analyses of Lyman band features 
arising from the $J=0$ and 1 rotational levels that appeared in FUSE spectra.  Here, 
the Galactic contributions were separable.

The remaining lower part of Table~\ref{tbl:col_dens_depl1} shows in alternating 
rows the logarithms of the column densities $N(X)$ (in units of ${\rm cm}^{-2}$) 
and depletions $D(X)$ for the elements $X$ considered in our study.  The 
uncertainties listed for $D(X)$ were computed from the sums in quadrature for the 
relative uncertainties in the element and the total hydrogen column densities.   We 
do not include uncertainties in the stellar reference abundances discussed in 
Section~\ref{sec:stellar_abundances} because they are themselves quite uncertain.  
Moreover, such errors in the reference abundances are irrelevant for the 
correlations of depletions discussed in Section~\ref{sec:correl} or the differential 
changes in the individual element depletions with respect to each other (as a 
function of our generalized depletion parameter), both of which are important 
themes in this investigation.

\section{DEPLETION RELATIONSHIPS}\label{sec:depl_relationships}

\begin{figure}
\plotone{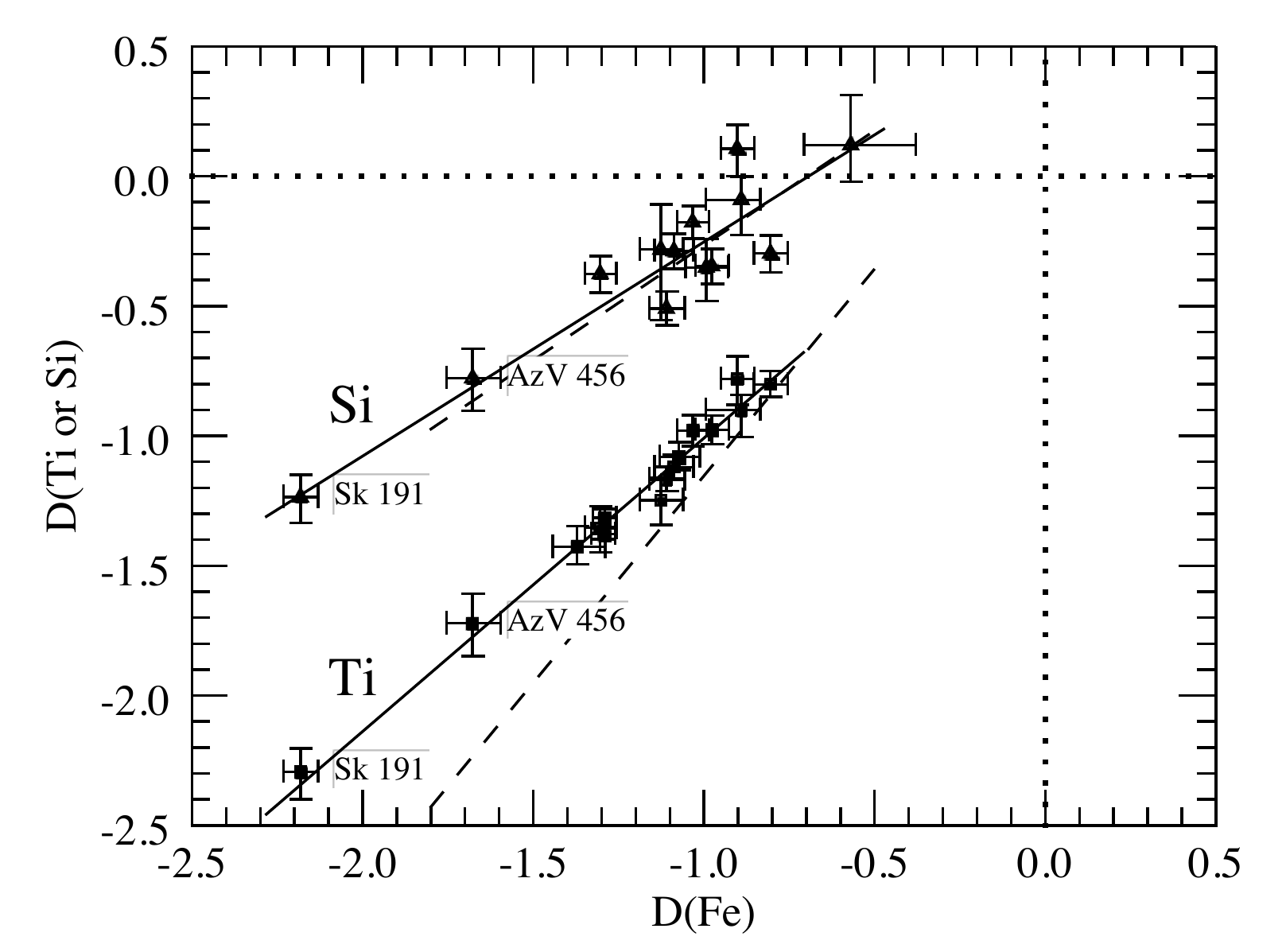}
\caption{Depletions $D$(Si) (triangles) and $D$(Ti) (squares) as a function of 
$D$(Fe) for different stars in our survey.   The stars AzV\,456 and Sk\,191 are 
identified; they represent special cases that are discussed in 
Section~\ref{sec:special_cases}.  The solid lines show the best fits to the data (from 
minimizing errors in both directions) and the dashed lines show the trends for 
depletions in the Milky Way using the coefficients given in Jenkins 
(2009).\label{fig:ti_si_vs_fe}}
\end{figure}

Figure~\ref{fig:ti_si_vs_fe} shows comparisons of $D$(Si) and $D$(Ti) for the 
corresponding values of $D$(Fe).   This figure shows that the depletions of Fe and Ti 
form a tight, linear sequence.  The results for $D$(Si) show more scatter, but this 
could be caused by larger uncertainties in the measurements.  The solid lines going 
through the points are best fits based on minimizing the errors in both directions, 
obtained through the use of the routine {\tt FITEXY} (Press et al. 2007).  If we use 
the depletion coefficients determined by Jenkins (2009), we can predict how the 
trends would look for gas in the Milky Way.  These relationships are shown by the 
dashed lines in the figure.  It is apparent that the behavior for $D$(Si) versus 
$D$(Fe) in the SMC is very similar to that in the Milky Way, but Ti shows a trend 
that is less steep in the SMC compared to that of our Galaxy.  (Note that similarities 
or differences in slope represent real propensities for elements to condense into 
solid form, while vertical displacements could arise simply from inaccuracies in our 
adopted reference abundances.)

\subsection{Pearson Correlation Coefficients between Elements}\label{sec:correl}

We expect that the depletions of all elements should be correlated with each other.  
Even so, the strengths of these correlations should be tested to see if the couplings 
between different combinations of elements show some variation.  This is not an 
easy task, since we must acknowledge that from one element to the next there are 
very different levels of accuracy for our measurements.  These variations will cause 
the depletions of some elements to show better or worse correlations with those of 
others.   Table~\ref{tbl:correl} shows the strengths of the correlations, expressed in 
terms of their Pearson correlation coefficients, which are listed in the upper right 
portion of the upper block of numbers.  The numbers of pairs for given 
combinations of elements are below the diagonal in this upper block.

As pointed out by Jenkins et al. (1986), the comparisons of depletions of elements 
can have spurious enhancements of their correlations because they share common 
errors in their denominators, $\log N({\rm H_{tot}})$.  Since we have estimates for 
the errors in $\log N({\rm H_{tot}})$ in our sample, we can correct the correlation 
coefficients to obtain more realistic values by applying the formulae given by 
Jenkins et al. (1986, Appendix B).  The numbers above the diagonal in the lower 
block of Table~\ref{tbl:correl} show these corrected correlation coefficients.  As 
expected, they are slightly lower than the raw correlation coefficients.

When a correlation of a small sample becomes marginal, it is important to know 
whether or not it is statistically significant, i.e., what are the chances that over a 
much larger population the correlation would vanish and the small sample drawn 
from this population showed a relationship simply by pure chance.  A
well-established procedure to test the significance of a positive correlation is to 
evaluate the quantity
\begin{equation}
t=r[(1-r^2)/(n-2)]^{-1/2}~,
\end{equation}
where $r$ is the observed correlation coefficient for $n$ pairs of samples.
The one-tail area of a Student's $t$-distribution applied to the value of $t$ gives the 
chance that the whole population correlation is less than or equal to zero.  The 
validity of this test relies on the proposition that the population is distributed much 
like a Gaussian distribution.  Since the results for the stars AzV\,456 and Sk\,191 
clearly seem far removed from those of the other stars, we eliminated these two 
cases when we evaluated the correlations and applied the tests.  The values listed 
below the diagonal of the lower block of numbers show the outcomes for one minus 
the one-tail area.

It is clear that a majority of combinations show that the positive correlations are 
real at or above the 95\% level of confidence.  Cases with lower confidence may 
arise simply from larger errors in the measurements or limited ranges over which 
measurements were possible (e.g., the weakness of the Mg~II lines and the 
saturations of the S~II lines limited their reliable ranges of measurement to only 
about 0.5~dex).  While this may be true, it is worthwhile to investigate whether or 
not some extreme differences in the apparent correlations could be real, which 
could be of significance in our interpretations of how the composition of dust 
particles might vary from one location to the next.  For instance, the corrected 
$r({\rm Mg,~Si})=0.596$ seems to differ from that of $r({\rm Mg,~Fe})=-0.050$.  
Does this signify that the depletions of Mg are actually more closely coupled to those 
of Si than to those of Fe, or could this be a statistical fluke?  If we examine the 95\% 
confidence intervals for the correlations $\rho$ that one would expect for 
populations much larger than our sample sizes, we conclude that $-0.25 \leq 
\rho({\rm Mg,~Si}) \leq +0.91$ and $-0.63 \leq \rho({\rm Mg,~Fe}) \leq +0.57$.  
Since there is a substantial overlap between these two intervals, we are unable to 
assert at a 95\% level of confidence that the correlations differ from each other, 
even though in fact they might do so.  In contrast to this outcome, another example 
points toward a real difference in correlations.  We identified two cases where we 
have 15 sample pairs that seemed to differ: $r({\rm Zn,~Fe})=0.438$ and $r({\rm 
Ti,~Fe})=0.992$, leading to $-0.12 \leq \rho({\rm Zn,~Fe}) \leq +0.75$ and 
$\rho({\rm Ti,~Fe}) \geq 0.95$, which are clearly incompatible with each other at 
the 95\% confidence level.  While this may be true, we note that $D$(Zn) does not 
correlate very well with any of the other elements, which may be just a consequence 
of the fact that the errors in measuring Zn depletions are large compared to the total 
range of the outcomes.

\begin{deluxetable*}{
c	
c	
c	
c	
c	
c	
c	
c	
c	
c	
}
\tablewidth{0pt}
\tablecaption{Pearson Correlation Coefficients\tablenotemark{a}\label{tbl:correl}}
\tablecolumns{10}
\tablehead{
\colhead{Element} & \colhead{D(Mg)} & \colhead{D(Si)} & \colhead{D(S)} & 
\colhead{D(Ti)} & \colhead{D(Cr)} & \colhead{D(Mn)} & \colhead{D(Fe)} & 
\colhead{D(Ni)} & \colhead{D(Zn)}
}
\startdata
D(Mg)&10&0.633&0.723&0.054&0.251&0.201&0.025&0.470&0.463\\
D(Si)&7&13&0.966&0.937&0.751&0.968&0.929&0.874&0.544\\
D(S)&6&10&10&0.908&0.708&0.939&0.897&0.914&0.765\\
D(Ti)&8&11&8&15&0.849&0.985&0.992&0.955&0.372\\
D(Cr)&9&8&6&10&13&0.894&0.868&0.845&0.484\\
D(Mn)&8&11&8&14&11&15&0.982&0.957&0.602\\
D(Fe)&10&13&10&15&13&15&18&0.934&0.481\\
D(Ni)&9& 9&8&11&12&11&14&14&0.702\\
D(Zn)&10&11&8&12&12&13&15&12&15\\ [10pt]
D(Mg)&\nodata&0.596&0.654&0.007&0.189&0.154&$-0.050$&0.427&0.428\\
D(Si)&0.921&\nodata&0.965&0.936&0.709&0.968&0.927&0.858&0.487\\
D(S)&0.921&0.997&\nodata&0.908&0.655&0.938&0.895&0.895&0.724\\
D(Ti)&0.506&0.963& 0.725&\nodata&0.840&0.985&0.992&0.952&0.339\\
D(Cr)&0.687&0.976&0.921&0.999&\nodata&0.886&0.857&0.832&0.432\\
D(Mn)&0.642&0.999&0.960&1.000&1.000&\nodata&0.982&0.954&0.571\\
D(Fe)&0.446&0.985&0.892&1.000&1.000&1.000&\nodata&0.929&0.438\\
D(Ni)&0.874&0.998&0.999&1.000&1.000&1.000&1.000&\nodata&0.675\\
D(Zn)&0.891&0.969&0.979&0.922&0.920&0.987&0.981&0.992&\nodata\\

\enddata
\tablenotetext{a}{{\it Upper half:\/} upper right portion: raw correlation 
coefficients, lower left portion: numbers of pairs.  {\it Lower half:\/} upper right 
portion: corrected correlation coefficients, lower left portion: the probabilities that 
the population correlations are greater than zero, given the values of the corrected 
correlation coefficients and sample sizes (but after omitting data for AzV\,456 and 
Sk\,191).}
\end{deluxetable*}

\subsection{Definition of a Generalized Depletion Parameter $F_*({\rm 
SMC})$}\label{sec:definition_F*}

Our principal objective is to describe the element depletions for the SMC gases 
within a simple framework that follows what Jenkins (2009) did for the local region 
of our Galaxy.  He invented a parameter that he called $F_*$ that characterized the 
general level of depletion for a sight line, and then was able to state that the 
depletion $[X_{\rm gas}/{\rm H}]$ of any element $X$ could be represented to good 
accuracy in the form
\begin{equation}\label{eqn:basic}
[X_{\rm gas}/{\rm H}]=B_X+A_X(F_*-z_X)~,
\end{equation}
where the coefficients $B_X$ and $A_X$ were unique for each element $X$, which he 
obtained from the best fits of the depletion data to the parameter $F_*$.  While the 
scale of $F_*$ was arbitrary, it was designed such that most sight lines fell within the 
range $0.0 \leq F_* \leq 1.0$.  The offset $z_X$ for the origin of $F_*$ creates a zero 
covariance in the uncertainties of $B_X$ and $A_X$ when they are evaluated by the 
best linear fits to the data.

The difference in depletion behaviors between Ti and Fe for the Milky Way and the 
SMC shown in Fig.~\ref{fig:ti_si_vs_fe} indicates that a definition of $F_*$ based on 
the depletions of different elements in the Milky Way is no longer appropriate.  For 
this reason, we must now define a new measure of $F_*$ that can be applied to gas 
in the SMC, which we will designate as $F_*({\rm SMC})$.  For the SMC, we do not 
have as broad a selection of elements and target stars to establish in an independent 
and self-consistent manner a depletion scale as Jenkins (2009) was able to do.  
Instead, we focus chiefly on the depletions of Fe, which was measured for every star 
in our SMC survey.  At the same time, it is desirable to design this new depletion 
parameter such that it is similar to the one defined earlier for the Milky Way, which 
we now call $F_*({\rm MW})$.  In order to do so, we have chosen to link the two 
scales through the depletion of Fe, which through a simple transformation of 
Eq.~\ref{eqn:basic} leads us to our proposed definition for any particular sightline,
\begin{equation}\label{eqn:F*(SMC)}
F_*({\rm SMC})={D({\rm Fe})-B_{\rm Fe}\over A_{\rm Fe}}+z_{\rm Fe}~,
\end{equation}
where we adopt the Milky Way values $B_{\rm Fe}= -1.51$, $A_{\rm Fe} = -1.28$, 
and $z_{\rm Fe} = 0.437$.

While the measurement uncertainties of $D({\rm Fe})$ are small, we can increase 
the accuracy of $ F_*({\rm SMC})$ if we also consider the depletions of two other 
elements, Ti and Mn, which are also well determined.  After evaluating the best fits 
of the depletions of these two elements relative to that of Fe [e.g., see 
Fig.~\ref{fig:ti_si_vs_fe} that shows $D({\rm Ti})$ vs $D({\rm Fe})$], for each 
individual sight line we use these best-fit relations to transform $D({\rm Ti})$ and 
$D({\rm Mn})$ to their equivalent Fe depletions and then evaluate a weighted 
average to optimize the equivalent value of the Fe depletion for determining 
$F_*({\rm SMC})$ through the application of Eq.~\ref{eqn:F*(SMC)}.  As with the 
measurements of column densities discussed in 
Section~\ref{sec:observing_strategy}, the weights are proportional to 
$1/\sigma_i^2$ and the final uncertainty in $F_*({\rm SMC})$ equals $\left(\sum_i 
\sigma_i^{-2}\right)^{-0.5}$.  Table~\ref{tbl:fstar} lists the outcomes for all of the 
stars in our survey.

\begin{deluxetable}{
c	
c	
}
\tablewidth{0pt}
\tablecaption{$F_*$(SMC)\label{tbl:fstar}}
\tablehead{
\colhead{Star} & \colhead{$F_*$(SMC)}
}
\startdata
AzV\,18&\phs $ 0.29\pm 0.01$\\
AzV\,26&\phs $ 0.14\pm 0.02$\\
AzV\,47&$-0.11\pm 0.03$\\ 
AzV\,78&\phs $ 0.09\pm 0.03$\\
AzV\,80&\phs $ 0.27\pm 0.02$\\
AzV\,95&\phs $ 0.02\pm 0.02$\\ 
AzV\,104&\phs $ 0.03\pm 0.05$\\
AzV\,207&\phs $ 0.14\pm 0.03$\\ 
AzV\,216&\phs $ 0.10\pm 0.03$\\ 
AzV\,229&\phs $ 0.03\pm 0.02$\\
AzV\,242&\phs $ 0.11\pm 0.02$\\
AzV\,321&$-0.05\pm 0.04$\\ 
AzV\,332&$-0.35\pm 0.10$\\
AzV\,388&\phs $ 0.27\pm 0.03$\\
AzV\,456&\phs $ 0.53\pm 0.04$\\
AzV\,476&\phs $ 0.33\pm 0.03$\\
Sk\,190&$-0.06\pm 0.03$\\
Sk\,191&\phs $ 0.94\pm 0.03$\\

\enddata
\end{deluxetable}
\newpage
\subsection{How Element Depletions Scale with $F_*({\rm 
SMC})$}\label{sec:trends_F*}

\begin{figure*}
\plotone{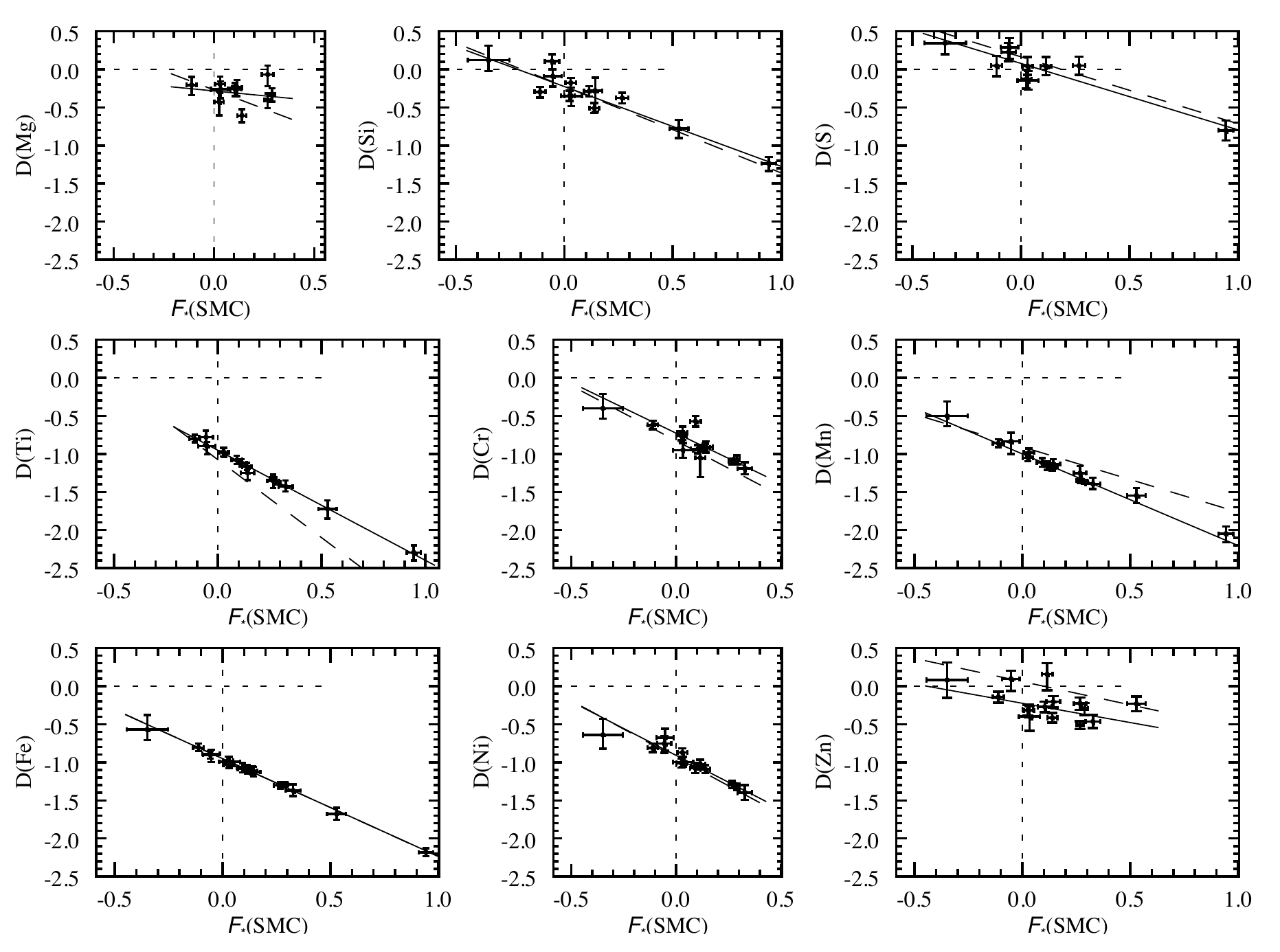}
\caption{Element depletions vs. $F_*$(SMC).  Solid lines show the best fits, with the 
coefficients given in Table~\ref{tbl:coeffs}.  Dashed lines represent fits that were 
obtained for our Galaxy.  By design, the fits for $D({\rm Fe})$ for our Galaxy and the 
SMC coincide with each other.  The points located at $F_*({\rm SMC})\approx 0.5$ 
and 0.9 represent AzV\,456 and Sk\,191, respectively.\label{fig:fstar_plots}}
\end{figure*}

For each element $X$, we determined the best fit of the depletions to the $F_*({\rm 
SMC})$ parameter, again by minimizing errors in both the dependent and 
independent quantities using the linear fit routine {\tt FITEXY}.  The outcomes for 
$A_X$, $B_X$, and $z_X$ are shown in the right-hand half of Table~\ref{tbl:coeffs}, 
which can be compared to the same set of coefficients that apply to the local 
$F_*({\rm MW})$ that are listed in the left-hand side of the table.  
Figure~\ref{fig:fstar_plots} shows plots that exhibit the depletion measurements 
versus $F_*({\rm SMC})$ along with the fits constructed from Eq.~\ref{eqn:basic}.  
It is important to emphasize that any deviations in the stellar reference abundances 
from our adopted $[X/{\rm H}]=-0.65$, as indicated by the (somewhat uncertain) 
numbers listed in Column~6 of Table~\ref{tbl:trans}, will be reflected in the form 
of vertical offsets in the trends, as defined by the $B_X$ parameters.  The $A_X$ 
slope parameters are immune to errors in these reference abundances.

\begin{deluxetable*}{
c	
c	
c	
c	
c	
c	
c	
c	
c	
c	
c	
}
\tablewidth{0pt}
\tablecaption{Fit Coefficients for $F_*$(MW) and $F_*$(SMC)\label{tbl:coeffs}}
\tablecolumns{11}
\tablehead{
\colhead{} & \multicolumn{3}{c}{Milky Way (MW)} & \colhead{~} & 
\multicolumn{6}{c}{Small Magellanic Cloud (SMC)}\\
\cline{2-4} \cline{6-11}
\colhead{$X$} & \colhead{$A_X$} & \colhead{$B_X$} & \colhead{$z_X$} & 
\colhead{~} & \colhead{$A_X$} &
\colhead{$B_X$\tablenotemark{a}} & \colhead{$z_X$} & \colhead{$\chi^2$} & 
\colhead{$\nu$} & \colhead{$P$\tablenotemark{b}}
}
\startdata
Mg&$-1.00\pm 0.04$&$-0.80\pm 0.02$& 0.531&&$-0.25\pm 0.26$&$-0.33\pm 
0.03$& 0.162&22.1& 8&0.005\\
Si&$-1.14\pm 0.06$&$-0.57\pm 0.03$& 0.305&&$-1.05\pm 0.09$&$-0.36\pm 
0.02$& 0.129&26.7&11&0.005\\
S&$-0.88\pm 0.28$&$-0.09\pm 0.04$& 0.290&&$-0.87\pm 0.14$&$-0.02\pm 
0.04$& 0.106&11.1& 8&0.196\\
Ti&$-2.05\pm 0.06$&$-1.96\pm 0.03$& 0.430&&$-1.45\pm 0.09$&$-1.23\pm 
0.02$& 0.189& 2.3&13&0.999\\
Cr&$-1.45\pm 0.06$&$-1.51\pm 0.05$& 0.470&&$-1.33\pm 0.16$&$-0.93\pm 
0.02$& 0.155&18.7&11&0.066\\
Mn&$-0.86\pm 0.04$&$-1.35\pm 0.03$& 0.520&&$-1.20\pm 0.09$&$-1.24\pm 
0.02$& 0.196& 5.1&13&0.973\\
Fe&$-1.28\pm 0.04$&$-1.51\pm 0.03$& 0.437&&$-1.28\pm 0.07$&$-1.18\pm 
0.02$& 0.181& 1.8&16&1.000\\
Ni&$-1.49\pm 0.06$&$-1.83\pm 0.04$& 0.599&&$-1.41\pm 0.14$&$-1.11\pm 
0.02$& 0.141& 7.0&12&0.857\\
Zn&$-0.61\pm 0.07$&$-0.28\pm 0.05$& 0.555&&$-0.51\pm 0.14$&$-0.31\pm 
0.02$& 0.168&44.8&13&0.000\\

\enddata
\tablenotetext{a}{The listed uncertainties represent only the formal ones in the fits 
to Eq.~\protect\ref{eqn:basic} without any consideration of the possible real 
deviations from the stellar reference abundances listed in Column~5 of 
Table~\protect\ref{tbl:trans}.}
\tablenotetext{b}{The probability of obtaining a worse fit to the linear trend, given 
the $\chi^2$ values and their respective degrees of freedom $\nu$, which are 
driven by our estimates for the individual errors in both directions.  A low value 
indicates that either we have underestimated the errors or that real depletions 
deviate from a simple linear relationship with $F_*({\rm SMC})$.}
\end{deluxetable*}

Figure~\ref{fig:fstar_plots} shows that the fits for Mg, Ti, and Mn for the Milky Way 
(dashed lines) differ from those for the SMC.  While there seems to be a vertical 
offset between the two for Zn, one could discount this as arising either from an error 
in the adopted SMC stellar abundance for Zn or the more complex nucleosynthesis 
origins for this element (note that the slopes $A_{\rm Zn}$ are nearly identical to 
each other to within their uncertainties).  The remaining elements, Si, S, Cr, and Ni, 
seem to have depletion sequences in the SMC that are very similar to those seen in 
the Milky Way (the two relationships for Fe are identical by construction).  Vladilo 
(2002) found that Mn seemed to exhibit stronger depletions in the SMC relative to 
the standard depletion patterns for other elements, which is consistent with our 
finding that $A_{\rm Mn}$ for the SMC indicates a more negative slope than its 
counterpart for the Milky Way.  The weaker Si depletions relative to the Galactic 
trends are not duplicated by our findings, where we determined that $A_{\rm Si}$ 
for the SMC is about identical to that for the Milky Way.
\newpage
\subsection{The Special Cases of AzV\,456 and Sk\,191}\label{sec:special_cases}

There are two stars that exhibited depletion outcomes that appeared to be markedly 
stronger than those for other stars in the survey.  One of them, AzV\,456, is the most 
highly reddened star in our sample (see Table~\ref{tbl:target_stars}), exhibits 
relatively strong absorptions by diffuse interstellar bands (Welty et al. 2006 ; Cox et 
al. 2007), shows absorption features arising from C$_2$, C$_3$, CN and CH (Welty et 
al. 2013), has a large fraction of hydrogen in molecular form (see 
Table~\ref{tbl:col_dens_depl1}) and, unlike most other stars in the SMC, exhibits a 
UV extinction curve that is similar to that of our Galaxy (Gordon \& Clayton 1998 ; 
Cartledge et al. 2005 ; Cox et al. 2007).  From their analysis of the rotational 
excitations of C$_2$ and H$_2$, Welty et al. (2013) found a low kinetic temperature 
and a moderately high density for the absorbing cloud in front of AzV\,456, although 
the thermal pressure derived from the C~I fine-structure excitation seems lower 
than $nkT$ indicated by the molecules (Welty et al. 2016).  For these reasons, it is 
no surprise that this star is well removed from the others in the depletion plots 
shown in Figs.~\ref{fig:ti_si_vs_fe} and \ref{fig:fstar_plots}.  We note that that its 
depletions for different elements seem to follow extrapolations of the sequences 
with $F_*$(SMC) for the other stars in our survey.

The other star, Sk\,191, shows even more depletion than AzV\,456.  This star is 
located in the southeastern ``wing" of the SMC, but unlike AzV\,456 the sight line 
toward this star does not exhibit an especially large reddening or high H$_2$ 
fraction.  This star is the furthest removed from the others and is close to the 
Magellanic Bridge (MB), a region containing gas and stars that spans the SMC and 
LMC.  Dynamical simulations by Gardiner \& Noguchi (1996) and Hammer et al. 
(2015) suggest that the MB was created by a tidal interaction when the Large and 
Small Magellanic Clouds collided some 200$-$300 Myr ago. Several stars embedded 
in the MB midway between the SMC and LMC show abundances even lower than 
those in the SMC (Rolleston et al. 1999 ; Lee et al. 2005 ; Dufton et al. 2008), and the 
results of Lehner et al. (2001); (2008) indicate that the same holds for gas in this 
region.

A legitimate question is whether or not the environment in front of Sk\,191 likewise 
has an intrinsically lower total abundance of heavy elements compared to the 
material in front of the other stars in the SMC (Welty \& Crowther 2010).   A 
revealing indication is that stars closest to Sk\,191 investigated by Lee et al. (2005) 
seem to have abundances that more closely match those of stars within the SMC 
proper, and the same holds for Sk\,191 itself (Trundle et al. 2004).  Moreover, the 
spectrum of Sk\,191 is one of the very few cases where we have a marginal 
detection of interstellar O~I absorption at 1356\,\AA\ that lines up perfectly at the 
correct velocity.  Our measurement of the equivalent width of this weak feature is 
$9.1\pm 7.6\,$m\AA, which leads to a value $\log N({\rm O~I})=17.68$.  If one 
imagines that the depletion of O is almost zero, we would expect to find $\log 
N({\rm O~I}) = 17.66$ for $\log N({\rm H}) =  21.62$ and [M/H] equal our adopted 
value for the SMC of $-0.65$.  If this detection is real and its magnitude is 
approximately correct, we would conclude that the intrinsic abundances of the gas 
in front of this star should not be appreciably lower than elsewhere in the SMC.  In 
short, various considerations favor the idea that extraordinarily strong depletions 
onto grains are responsible for the low abundances toward this star.

\subsection{Sulfur: A Troublesome Element}\label{sec:sulfur}

The title of this subsection is identical to one that appeared in Jenkins (2009).  In 
studies of S~II absorption toward stars in the Milky Way, most of the sight lines that 
satisfied a criterion that $\log N({\rm H_{tot}}) > 19.5$ (to eliminate the need for 
ionization corrections) also had enough ionized sulfur to make all three lines of the 
triplet completely saturated and hence unusable.  Even for the few cases where 
acceptable results for S could be obtained, there was also the concern that some of 
the S~II could reside within the H~II regions surrounding the target star, since the 
ionization potential of S$^+$ is high (23.4\,eV).  There were a number of examples 
in the study by Jenkins (2009) that indicated the presence of apparent super-solar 
abundances of S if one assumed that all of the S was in an H~I region.

For the SMC, the overall lower metallicity of the gas gives us a larger margin to 
observe absorption features that are not totally saturated and yet have enough total 
gas in the sight line so that ionization corrections and the contributions from H~II 
regions are not substantial.  The stellar temperatures of most of the stars in our 
survey are high enough that appreciable fractional concentrations of S$^+$ should 
only be expected in the outer few percent of a line through the H~II region (Sarazin 
1977 ; Evans \& Dopita 1985 ; Evans 1991).  However, it is possible that some of our 
sight lines may accidentally penetrate the H~II regions of foreground B-type stars, 
where there are far fewer photons that can convert S to the doubly ionized state 
because their fluxes are strongly attenuated at energies that are above the 
ionization potential of neutral He at 24.6\,eV (Vallerga \& Welsh 1995 ; Cassinelli et 
al. 1996), which nearly coincides with that of S$^+$.   However, such stars can still 
fully ionize the accompanying hydrogen.

The results in Table~\ref{tbl:col_dens_depl1} indicate that for three of our stars, 
AzV\,321, AzV\,332, and Sk\,190, the interstellar S abundances appear to be 
approximately twice as large as the stellar abundance standard that we adopted, i.e., 
$D({\rm S})\approx +0.3$.  Values of $N({\rm H_{tot}})$ for these three stars are 
lower than for all of the others in our sample, they have negative values for 
$F_*$(SMC), and there is weak evidence that both $D$(Si) and $D$(Zn) may be 
positive, but by amounts that are not as significant as our findings for S.  Deviations 
in the measured stellar abundances of S from our adopted reference standard 
appear to be small, as indicated by the entry for this element in Column~6 of 
Table~\ref{tbl:trans}.  Thus, as with the Milky Way, the results for S in the SMC 
present the puzzling outcomes that interstellar S abundances occasionally exceed 
those for young stars in the same general environment.  However, the possibility 
that the accompanying Si and Zn might also have higher than expected 
concentrations may indicate that local abundance anomalies may be responsible for 
these deviations (the separation in the sky for AzV\,321 and AzV\,332 is only 
$155\arcsec$). 
\begin{figure*}[t]
\plotone{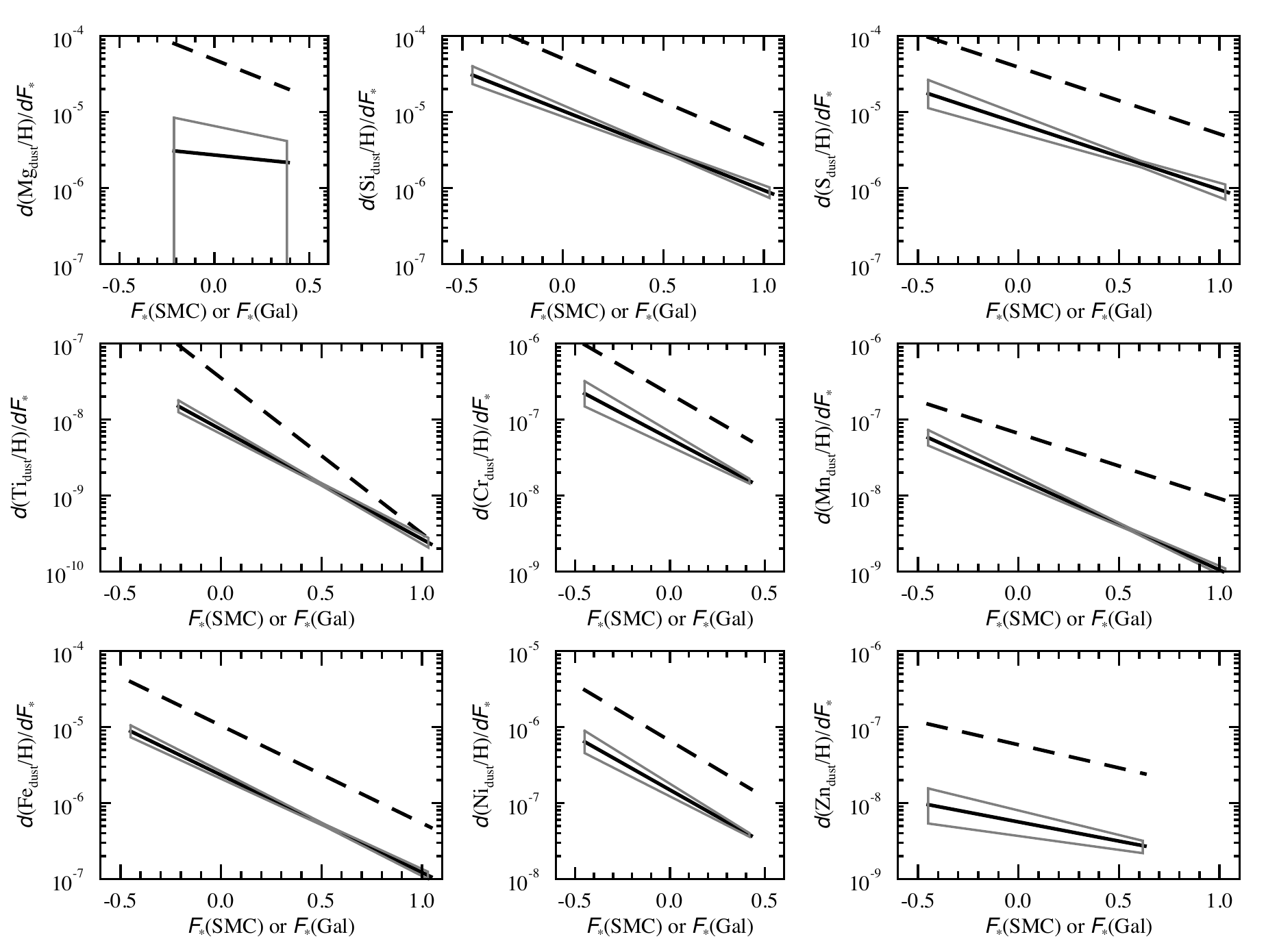}
\caption{{\it solid lines:\/} Differential consumptions of elements by number 
(relative to hydrogen) by dust grains for small changes in $F_*({\rm SMC})$ plotted 
as a function of $F_*({\rm SMC})$.  The surrounding uncertainty envelopes 
representing the $\pm 1\sigma$ deviations in $A_X$ and $B_X$ are shown by light 
lines.  {\it Dashed lines:\/} The same for dust consumption in the Milky Way (as a 
function of $F_*({\rm MW})$).  The large vertical displacements between the two 
lines reflect that fact that the SMC has a 0.65\,dex lower metallicity than our Galaxy, 
so changes in depletion in the SMC result in fewer atoms condensing onto the grains. 
\label{fig:dust_buildup}}
\end{figure*}

\section{BUILDUP OF ELEMENTS IN DUST GRAINS}\label{sec:buildup}

If we consider the amount of any particular element in dust, relative to the amount 
of hydrogen,
\begin{equation}\label{eqn:dust}
(X_{\rm dust}/{\rm H})=(X/{\rm H})_{\rm stellar}(1-10^{D_X})~,
\end{equation}
combine it with Eq.~\ref{eqn:basic}, and differentiate with respect to $F_*({\rm 
SMC})$, we can monitor the rates of condensation of elements onto grains as 
$F_*({\rm SMC})$ increases using the equation
\begin{eqnarray}\label{differential_grain_comp}
{d(X_{\rm dust}/{\rm H})\over dF_*({\rm SMC})}&=&-(\ln 10)(X/{\rm H})_{\rm 
stellar} A_X10^{A_X(F_*({\rm SMC})-z_X)+B_X}\nonumber\\
&=&-(\ln 10)A_X(X_{\rm gas}/{\rm H})_{F_*({\rm SMC})}
\end{eqnarray}
This differential analysis has the advantage of being independent of the adopted 
stellar abundances and their associated uncertainties.  It requires a knowledge of 
only the slope parameter $A_X$ and the (linear) expression for the relative 
abundance of the element $X$ with respect to ${\rm H_{tot}}$ at a particular value 
of $F_*({\rm SMC})$.  Figure~\ref{fig:dust_buildup} shows the trends for the 
consumptions of atoms in both the SMC and the Milky Way.
\newpage
\section{DISCUSSION}\label{sec:discussion}
\subsection{General Remarks}\label{sec:general}

The depletion trends that we presented in Section~\ref{sec:trends_F*} indicate that 
most elements deplete together in a manner similar to that of the Milky Way (e.g., 
compare the solid and dashed lines in Fig.~\ref{fig:fstar_plots}).  Two exceptions 
that we found to be significant are the elements Ti and Mn, as indicated by our 
finding that the outcomes for $A_{\rm Ti}$ and $A_{\rm Mn}$ in the SMC differ from 
their counterparts in the Milky Way by more than their uncertainties (see 
Table~\ref{tbl:coeffs}).  (Apparent differences in the listed $B_X$ values are of no 
fundamental significance, since they could be influenced by possible deviations in 
the true total abundances from our adopted reference abundances and also the 
differences in the $z_X$ values.)  Mg does not appear to deplete as rapidly in the 
SMC as it does in the Milky Way, but we have fewer measurements (because they 
were more difficult to perform), and the scatter in the outcomes is large, as 
indicated by the low value for $P$ in the last column of Table~\ref{tbl:coeffs}.  The 
scatter in the Zn depletions is large, although the worst fluctuations occur for 
measurements with the largest uncertainties.

\subsection{A Comparison with the SMC Depletions of Tchernyshyov et al. 
(2015)}\label{sec:T15}

Tchernyshyov et al. (2015) have conducted a depletion study of gas within both the 
SMC and LMC, based on UV spectroscopic data from FUSE and the COS on the HST.  
Their objective was similar to ours: they made comparisons of depletions of 
different elements and derived relationships using the same basic formalism as 
ours, except that they did not use an offset $z_X$ to eliminate error covariances in 
the slope and overall depletion parameters.  For elements and target stars that are 
common to our two surveys, the results for column densities agree with each other 
to within about 0.1\,dex, except for Si~II and Zn~II toward AzV\,456.  Their 
definition for their generalized depletion strength, $F^*$, differs slightly from our 
$F_*({\rm SMC})$.  If we anchor the two systems to the depletion trends of Fe, an 
element with the best measurement accuracies in both surveys, we derive a 
relationship  $F^*=1.13F_*({\rm SMC})+0.334$.  To convert from their 
parameters\footnote{We append (T15) to these parameters to avoid confusion with 
our parameters.} $A_X({\rm T15})$ and  $\delta(X)_0({\rm T15})$ to ours, we 
apply the transformations
\begin{equation}\label{eqn:TA}
A_X=1.13A_X({\rm T15})
\end{equation}
and
\begin{equation}\label{eqn:TB}
B_X=A_X({\rm T15})(1.13z_X+0.334)+\delta(X)_0({\rm T15})~.
\end{equation}
Columns~2$-$5 of Table~\ref{tbl:comparison} show a comparison of their findings 
to ours, after converting their parameters to our system through Eqs.~\ref{eqn:TA} 
and \ref{eqn:TB}.  Our values and theirs overlap to within the uncertainty ranges of 
both determinations, except for $B_{\rm Si}$ and $A_{\rm Zn}$, where for each case 
our lower bound approximately matched their upper bound. This table also shows 
their outcomes for the element P, which was not included in our survey.

\begin{deluxetable*}{
l	
c	
c	
c	
c	
c	
c	
c	
c	
}
\tablewidth{0pt}
\tablecaption{Parameter Comparisons\tablenotemark{a}\label{tbl:comparison}}
\tablecolumns{9}
\tablehead{
\colhead{} & \multicolumn{2}{c}{SMC Values from} &
\colhead{} & \colhead{} & \colhead{} & \colhead{} & 
\multicolumn{2}{c}{DLA Values from}\\
\colhead{} & \multicolumn{2}{c}{Tchernyshyov et al. (2015)} &
\colhead{} & \multicolumn{2}{c}{SMC Values from} & \colhead{} & 
\multicolumn{2}{c}{De Cia et al. (2016)}\\
\colhead{} & \multicolumn{2}{c}{(using Eqs.~\protect\ref{eqn:TA} and 
\protect\ref{eqn:TB})} & \colhead{~} & \multicolumn{2}{c}{this 
Paper\tablenotemark{b}} & \colhead{~} & 
\multicolumn{2}{c}{(Using Eqs.~\protect\ref{eqn:DA} and \protect\ref{eqn:DB})} 
\\
\cline{2-3} \cline{5-6} \cline{8-9}
\colhead{Element} & \colhead{$A_X$} & \colhead{$B_X$}&
& \colhead{$A_X$} & \colhead{$B_X$} &
& \colhead{$A_X$} & \colhead{$B_X$} \\
\colhead{(1)} & \colhead{(2)} & \colhead{(3)} & & \colhead{(4)} &
\colhead{(5)} & & \colhead{(6)} & \colhead{(7)}
} 
\startdata
O&\nodata&\nodata&&\nodata&\nodata&&$-0.15\pm0.04$&$-0.15\pm 
0.13$\tablenotemark{c}\\
Mg&\nodata&\nodata&&$-0.25\pm 0.26$&$-0.33\pm 0.03$&&$-0.62\pm 
0.05$&$-0.58\pm 0.07$\\
Si&$-1.16\pm 0.17$&$-0.50\pm0.11$&&$-1.05\pm0.09$&$-0.36\pm 0.02$&&$-
0.64\pm 0.06$&$-0.58\pm 0.06$\\
P&$-0.94\pm 0.08$&$-0.32\pm 
0.05\tablenotemark{c,d}$&&\nodata&\nodata&&$-0.10\pm 0.07$&$-0.08\pm 
0.09$\tablenotemark{c}\\
S&\nodata&\nodata&&$-0.87\pm 0.14$&$-0.02\pm0.04$&&$-0.28\pm 0.08$&$-
0.28\pm 0.07$\\
Cr&$-1.10\pm 0.12$&$-0.85\pm 0.08$&&$-1.33\pm 0.16$&$-0.93\pm 0.02$&&$-
1.34\pm 0.10$&$-1.04\pm 0.05$\\
Mn&\nodata&\nodata&&$-1.20\pm0.09$&$-1.24\pm0.02$&&$-0.97\pm 0.04$&$-
0.86\pm 0.05$\\
Fe\tablenotemark{e}&$-1.28\pm 0.08$&$-1.18\pm 0.06$&&$-1.28\pm 0.07$&$-
1.18\pm 0.02$&&$-1.28\pm0.04$&$-1.18\pm 0.05$\\
Zn&$-0.70\pm 0.07$&$-0.36\pm0.07$&&$-0.51\pm 0.14$&$-0.31\pm 0.02$&&$-
0.27\pm 0.03$&$-0.25\pm 0.03$\\

\enddata
\tablenotetext{a}{Eqs.~\protect\ref{eqn:TA}, \protect\ref{eqn:TB}, 
\protect\ref{eqn:DA}, and \protect\ref{eqn:DB} are used to convert the coefficients 
for the relationships found by the two other investigators into our $A_X$ and $B_X$ 
parameters that are linked to $F_*({\rm SMC})$.}
\tablenotetext{b}{Copied from Table~\protect\ref{tbl:coeffs}  for ease of 
comparison.}
\tablenotetext{c}{This value was derived by arbitrarily setting $z_X$ equal to 0.15, 
which is representative of $z_X$ for the other elements.} 
\tablenotetext{d}{In their study of SMC and LMC depletions, Tchernyshyov et al. 
(2015) mistakenly swapped the SMC and LMC reference abundances for P.  We have 
applied an upward correction for $B_{\rm P}$ of  0.4\,dex to overcome this error.} 
\tablenotetext{e}{The pairs of values of $A_X$ and $B_X$ are identical by design 
since the depletion relationships of Fe were used to tie all of the parameter systems 
together.}
\end{deluxetable*}

\subsection{Relating our Results to DLA Depletions}\label{sec:relatingDLA}

While our investigation is intended to guide future interpretations of gas-phase 
abundances measured for DLA and sub-DLA systems, it is also useful to look back on 
some previous investigations of these systems in order to learn how well, or how 
poorly, our depletion patterns match them.  We discussed in 
Section~\ref{sec:motivation} some examples of puzzling deviations from the 
Galactic depletion patterns.  The few departures that we see between the SMC and 
the Milky Way indicate that some depletions do not behave in a universal fashion for 
all possible combinations of metal-bearing systems.

Ledoux et al. (2002) examined the trends of Si, Ti, Cr, and Mn abundances relative to 
Fe as a function of [Zn/Fe].  It is generally accepted that the depletions of Zn are 
weak and those of Fe are strong, so their depletion ratio can serve as a proxy for the 
relative generalized depletions.  They found that [Si/Fe] increases with increasing 
[Zn/Fe], which is qualitatively consistent with our conclusion that in the SMC 
$A_{\rm Si}>A_{\rm Fe}$ (i.e., Si depletes less rapidly than Fe; note that in the Milky 
Way the difference between the two may be slightly less).  Their values of [Cr/Fe] 
seem to be independent of [Zn/Fe], which agrees with our conclusion that $A_{\rm 
Cr}\approx A_{\rm Fe}$ in the SMC (Cr depletes more rapidly than Fe in the Milky 
Way).  However, they find substantial increases in [Mn/Fe] when [Zn/Fe] becomes 
larger, which seems inconsistent with our determination that $A_{\rm Mn}$ is not 
much different from $A_{\rm Fe}$ in the SMC.  The trend for Mn seems more 
consistent with the pattern in the Milky Way, where $A_{\rm Mn}$ shows a 
substantially shallower slope than that for $A_{\rm Fe}$. 

A recent, more comprehensive investigation of DLA element depletions has been 
carried out by De Cia et al. (2016).  They too used [Zn/Fe] as an indicator for 
depletion, but they also made use of the finding that [Zn/Fe] is correlated with 
[Zn/H] (Wolfe et al. 2005), which serves as an approximate indicator of overall 
metallicity.  After examining offsets in some relative abundances in DLAs, they 
determined corrections for the zero-depletion element abundances caused by 
nucleosynthesis effects in such systems, with some additional guidance from the 
trends seen in metal-poor stars in our Galaxy.  After applying these corrections, they 
derived for each element the coefficients $A_2$ and $B_2$ that gave a best fit for the 
depletions that satisfied the equation $D=A_2+B_2[{\rm Zn/Fe}]$.

In principle, we could compare our results to those of De Cia et al. (2016) using the 
parameter [Zn/Fe] as a common measure of depletion strengths.  However, our 
determination for the trend of [Zn/Fe] as a function of $F_*({\rm SMC})$ has a large 
uncertainty - much larger than for our measurements of $D({\rm Fe})$.  Hence, as 
we did for the results of Tchernyshyov et al. (2015) in Section~\ref{sec:T15}, we 
link our depletion sequence to that of De Cia et al. (2016) by using the depletions of 
Fe and find that their depletion parameter [Zn/Fe] can be linked to ours using the 
expression $[{\rm Zn/Fe}]=1.016F_*({\rm SMC})+0.744$.  We then find that the 
transformations to our coefficients from theirs with subscripts $2,X$ for various 
elements $X$ (see their Table~3) are given by
\begin{equation}\label{eqn:DA}
A_X=1.016B_{2,X}
\end{equation}
and
\begin{equation}\label{eqn:DB}
B_X=B_{2,X}(1.016z_X+0.744)+A_{2,X}~.
\end{equation}
As we did for the results of Tchernyshyov et al. (2015), we once again compare 
coefficients arising from the above two equations to those that we derived for the 
SMC, and we show the outcomes in Columns~4$-$7 of Table~\ref{tbl:comparison}.

In agreement with the results from Ledoux et al. (2002), De Cia et al. (2016) find 
that Mn depletes less rapidly than Fe, whereas we found that the difference between 
the two slopes was not significant.  The pronounced trends for [Mn/Fe] as a function 
of [Zn/Fe] in DLAs may signify the existence of either some chemical evolution 
effect that has not yet been recognized or a depletion sequence that more closely 
matches the behavior in our Galaxy.  The value for $A_{\rm Mg}$ in the DLAs seems 
to be midway between the Milky Way and SMC values for this slope parameter.  The  
slope for the depletion trend of Si is only half as large that for Fe, and the absolute 
value for $A_{\rm P}$ is quite small, in contrast to the findings for both the Milky 
Way and the SMC.

\subsection{A Possible Influence from the Abundance of Carbon}\label{sec:carbon}

In Section~\ref{sec:stellar_abundances} we adopted the viewpoint that all of the 
elements that we considered had abundances that were uniformly below the solar 
abundances by about $-0.65$\,dex, at least to within an envelope of apparent 
deviations of order 0.1 to 0.2\,dex.  It is natural to question why this dilute mixture 
of heavy elements (relative to H) with an otherwise identical composition should 
deplete onto dust differently in the SMC than what we found in our Galaxy.  The 
answer may lie in the abundance of the element carbon, which up to now we have 
not considered in this investigation.  Carbon is a key constituent in polycyclic 
aromatic hydrocarbons (PAHs).  Calculations by Weingartner \& Draine (1999) 
emphasize the probable importance for the binding of heavy elements onto very 
small PAHs.  If the balance in the SMC between PAHs (or other carbonaceous 
compounds) and silicates differs from that of our Galaxy, then the mix of substrates 
upon which additional atoms can condense will be different. Differences in the 
chemical affinities of various elements on existing grains could influence the growth 
and destruction rates for the outer portions of the grains. 

Determinations of stellar abundances in the SMC indicate that carbon may be more 
deficient than the other elements.  Measurements of the features of C~II and C~III 
in two main-sequence B-type stars in the SMC and an application of a non-LTE 
analysis by Hunter et al. (2005) yielded $\log ({\rm C/H}) + 12 = 7.42$, which is $-
1.01$\,dex below the solar abundance of carbon and $-0.36$\,dex below our 
average of $-0.65$\,dex that we adopted for all of the other elements.  Other good 
determinations yielded even lower values for the SMC stellar abundances, e.g., 7.30 
(Trundle et al. 2004 ; Dufton et al. 2005 ; Hunter et al. 2009) and 7.35 (Hunter et al. 
2007).  The outcomes for the C abundances are not surprising: stars that have 
$6.6\lesssim \log({\rm O/H})+12\lesssim 7.6$ (i.e., $-1.8 \lesssim [{\rm O/H}] 
\lesssim -0.8$) in the Milky Way halo likewise exhibit unusually low values of [C/O], 
i.e., $[{\rm C/O}]\approx -0.5$  (Akerman et al. 2004).  Moreover, the deviation of C 
below the reduced abundances of other elements in the SMC seems to be duplicated 
in other dwarf galaxies with low metallicities.  For instance, a study of nebular 
emission lines by Berg et al. (2016) found that when $\log ({\rm O/H})+12 \leq 
8.0$, [C/O] reverts to a mean value of $-0.36$, but with a large dispersion 
(0.25\,dex) from one system to the next.  The more strongly reduced carbon 
abundances seen in the SMC, Milky Way halo stars, and other metal-poor systems 
may reflect the increasing importance of a secondary production of carbon in low 
and intermediate mass stars as the overall abundances approach the solar value.

Entirely different kinds of observations seem to support the aforementioned carbon 
abundance trend: the emission from the PAH features in the mid-infrared appear to 
undergo an abrupt weakening for galactic systems when the oxygen abundances 
appear to be diminished to values $\log ({\rm O/H})+12 \lesssim 8.0$ (Wu et al. 
2006 ; Draine et al. 2007 ; Engelbracht et al. 2008).  In addition, if we accept the 
interpretation that the 2175\,\AA\ extinction bump is likely to be caused by large 
carbon compounds (Draine 2003), the lack of this feature in the extinction curves 
associated with QSO absorption-line systems (Pei et al. 1991 ; York et al. 2006 ; 
Budzynski \& Hewett 2011 ; Khare et al. 2012) helps to reinforce our perception 
about the apparent extra deficiency of carbon in systems with low metallicities.  
However, studies of DLAs with extremely low metallicities ($\log ({\rm 
O/H})+12\lesssim 7$) by Pettini et al. (2008), Penprase et al. (2010), and Cooke et 
al. (2011) revealed that when metal deficiencies become more extreme, the values 
of [C/O] begin to revert upward toward zero.

Much of what we said about distant, low-metallicity systems in general applies to 
the SMC.  With one exception that we discussed in Section~\ref{sec:special_cases}, 
more than half of the sight lines with A$_{\rm V} > 0.1$ through the SMC exhibit 
extinction curves that lack a detectable 2175\,\AA\ bump (Hagen et al. 2016).  Li \& 
Draine (2002) determined that PAH mid-IR emission from the interstellar medium 
of the SMC is relatively weak, which supports the interpretation that carbon 
compounds are deficient.  Sandstrom et al. (2010) performed a more 
comprehensive investigation of the PAH emission using the IRAC, MIPS and IRS 
instruments on the {\it Spitzer} facility.  They concluded that the fractional 
abundance of PAHs in the diffuse SMC material is low: the $q_{\rm PAH}$ 
index\footnote{This index for the concentration of PAHs is defined by Draine \& Li 
(2007) and Draine et al. (2007) as the mass fraction of dust that consists of PAH 
particles with less than $10^3$ carbon atoms.} averages about 0.6\%, and rises to 
$q_{\rm PAH} \sim 1-2\%$ in the dense molecular regions.  By comparison, 
$q_{\rm PAH} \approx 4.6\%$ in the Milky Way (Draine \& Li 2007).

One challenge to our suggestion that a deficiency of carbon compounds may be 
responsible for the differences between the SMC and Milky Way depletions is our 
finding that the results for AzV\,456 seem to fit well with the sequences for different 
elements that we found for other stars.  However, as we pointed out in 
Section~\ref{sec:special_cases}, the line of sight to this star shows an extinction 
curve that is unlike those of other SMC cases and closer to the behavior for Milky 
Way sight lines.  From a different perspective, a map shown by Sandstrom et al. 
(2010) indicates that the stars AzV\,18, AzV\, 26, and AzV\,47 are positioned in a 
region where $q_{PAH}\sim 1\%$, which is higher than other regions where most 
of our stars are located.  These stars do not seem to depart from the depletion 
sequences shown by the other ones.  Nevertheless, one could argue that the gas 
cloud with enhanced values of $q_{\rm PAH}$ could be situated behind these three 
stars.

\section{SUMMARY}\label{sec:summary}

Our principal objective was to investigate how rapidly different elements condense 
into the solid phase within the interstellar medium of the SMC.  Since the chemical 
evolution of the SMC is probably similar to other galactic systems with moderately 
low metallicities at redshifts $z \lesssim 3.5$, our depletion sequences could serve 
as a guide for understanding the removal of elements from the gas phase.  In turn, 
this information will aid in the determination of total element abundances in DLAs 
and sub-DLAs.  In addition, the SMC depletions help us to understand better the 
compositions of dust grains in the SMC, which can be compared to those in our 
Galaxy.

The nature and results of our investigation are as follows:
\begin{enumerate}
\item We conducted observations of the spectra over the wavelength range 1123$-
$2673\,\AA\ for 14 hot stars in the SMC using the E140M and E230M echelle modes 
of the STIS instrument on the HST during the Cycle~22 observing session.  We 
supplemented these data with existing UV spectra in the MAST archive for 3 other 
stars and column densities in the literature for one additional star.  We derived 
column densities for Mg~II, Si~II, S~II, Cr~II, Mn~II, Fe~II, Ni~II, and Zn~II, and 
we made use of Ti~II results published by Welty \& Crowther (2010).  We obtained 
atomic and molecular hydrogen column densities toward our stars from Welty et al. 
(2012).
\item For the total element abundance standards against which to compare the 
interstellar abundances, we have examined the findings reported in the literature 
for the abundances in young stars within the SMC.  For investigations that we 
deemed to be most reliable, we found that the dispersions in outcomes for most 
elements were on the order of 0.15\,dex, although we recognize that additional 
errors of an unknown systematic nature could exist for certain elements.  Except for 
carbon, the abundances of $\alpha$-process with respect to Fe-group elements 
seemed not to differ appreciably from the solar abundance ratios.   We found no 
compelling reason for adopting an SMC abundance pattern that was more complex 
than a uniform deficiency of $-0.65$\,dex below the solar abundances defined by 
Asplund et al. (2009).
\item We presented in Table~\ref{tbl:col_dens_depl1} column densities and the 
depletion outcomes defined in Eq.~\ref{eqn:depletion} for all measurements where 
the absorption features were strong enough to measure above the noise but not so 
strong that they were too saturated to yield useful results.  These restrictions 
precluded our being able to measure the abundances of B, C, N, O, P, Cu, Ge, Ga, and 
Kr, even though they had transitions within our wavelength coverage.
\item We found that most of the possible pairs of elements exhibited tight linear 
relationships with each other in their depletions.  In Section~\ref{sec:correl} we 
analyzed the strengths of the correlations and found that many of them (23 out of 
36) were significant at or above the 95\% confidence level.  The remaining cases 
showing weaker correlations could mostly be attributed to larger uncertainties in 
the measurements, reduced dynamic ranges, or small sample sizes.  We found that 
the loose correlation between the depletions of Zn and Fe appear to differ from the 
much tighter correlation for those of Ti and Fe at the 95\% confidence level, which 
suggests that there may be some variation in the Zn and Ti concentrations in dust in 
different regions.  Another possibility is that possibly different nucleosynthesis 
sources, combined with poor mixing in the SMC ISM, may explain the apparent 
variation in Zn depletions.
\item Following a scheme introduced by Jenkins (2009) for depletions in our 
Galaxy, we developed a unified description of depletions in terms of a generalized 
sight-line depletion parameter $F_*({\rm SMC})$ and two coefficients for each 
element that described a best-fit slope and offset in the linear relationship with this 
parameter (see Eq.~\ref{eqn:basic}).  We used Fe depletions to link the scale for 
$F_*({\rm SMC})$ to that of $F_*({\rm MW})$ in the Milky Way.  Values of 
$F_*({\rm SMC})$ for each star in our program are listed in Table~\ref{tbl:fstar}, 
and the coefficients $A_X$, $B_X$ and $z_X$ for each element $X$ are listed in the 
right-hand portion of Table~\ref{tbl:coeffs}.  From the $\chi^2$ values associated 
with the fits, we found that the elements Mg, Si, and Zn showed exceptionally low 
values for the probabilities of obtaining worse fits to the $F_*({\rm SMC})$ trends, 
indicating that either our measurement uncertainty estimates were too low or that 
there are true variations in relative abundances beyond those attributable to our 
regime of generalized depletions.
\item We identified two stars that were conspicuously different from the others in 
our survey.  The star AzV\,456 has an unusually high abundance of molecules in 
front of it and exhibits a UV extinction curve that is more similar to that of the Milky 
Way than the one typically found for matter in the SMC.  The depletions toward this 
star are strong, but they seem to lie on extrapolations of the trends seen for other 
stars.  The star Sk\,191 shows even more depletion than AzV\,456.  This star is 
located midway between the wing of the SMC and the Magellanic Bridge.  While one 
might be suspicious that the overall abundances of material near this star may be 
lower than usual for the SMC, the star itself shows element abundances close to 
others in the SMC.  Moreover, there is an indication from a marginal detection of the 
semi-forbidden O~I absorption feature at 1356\,\AA\ that the total ISM abundances 
are not depressed below those seen elsewhere in the SMC.
\item We examined differential changes in dust elemental compositions as a 
function of $F_*({\rm SMC})$ and found that, aside from the effect of a lower overall 
metal abundance, the general pattern of consumption of the different atoms was 
about the same as in the Milky Way, except for the few elements Mg, Ti, and Mn, 
where their respective values of $A_X$ for the SMC differed from their counterparts 
in the Milky Way.  One important feature of this differential measurement is that the 
outcome is not dependent on knowing the stellar reference abundance.
\item Our study overlaps the findings of Tchernyshyov et al. (2015) for the elements 
Si, Cr, Fe, and Zn.  We converted their coefficients to our system and found that, with 
the exceptions of $B_{\rm Si}$ and $A_{\rm Zn}$, the values overlap to within the 
uncertainties of both determinations.  We did not include phosphorous in our 
survey, so we converted the values of $A_{\rm P}$ and $\delta({\rm P})_0$ of 
Tchernyshyov et al. (2015) to our $A_{\rm P}$ and $B_{\rm P}$ for an assumed 
value $z_{\rm P}=0.15$ (see Table~\ref{tbl:comparison}).
\item The relative amount of dust in the form of PAHs in the SMC is significantly 
lower than in the Milky Way.  If the chemical affinities of some of our elements on 
PAHs differ from those with other types of dust, such as silicates, we might have an 
explanation for why some of our depletion trends are either steeper or shallower 
than in the Milky Way.  Other possible influences might include a generally more 
intense UV radiation field, lower H$_2$ fractions, or greater H$_2$ rotational 
excitations (Tumlinson et al. 2002).  Otherwise, it is difficult to understand why the 
dilute mix (relative to hydrogen) of elements would behave differently than what 
we observe locally.
\end{enumerate}

\acknowledgments
Support for the HST observing program nr. 13778 was provided by NASA through a 
grant from the Space Telescope Science Institute (STScI), which is operated by the 
Associations of Universities for Research in Astronomy, Incorporated, under NASA 
contract NAS5-26555.  Some of the data presented in this paper were obtained from 
the {\it Mikulski Archive for Space Telescopes\/} (MAST) maintained by the STScI.  
G.W. thanks the University of Washington for providing office space and a desktop 
computer to emeritus professors.  The authors thank D.~E.~Welty and 
B.~T.~Draine for useful discussions.  Dr.~Welty also furnished the error estimates 
for the hydrogen column densities, and S.~Woosley gave us advice on the 
nucleosynthesis of Zn.  We thank the anonymous referee for undergoing a careful 
reading of our paper and providing thoughtful comments.

\facility{HST (STIS), HST (GHRS)}
\software{FITEXY (Press et al. 2007)}


\begin{references}


\reference{7332} Akerman, C. J., Carigi, L., Nissen, P. E., Pettini, M., \& Asplund, M. 
2004, A\&A, 414, 931
\reference{7052} Asplund, M., Grevesse, N., Sauval, A. J., \& Scott, P. 2009, ARA\&A, 
47, 481
\reference{8770} Azzopardi, M., \& Vigneau, J. 1982, A\&AS, 50, 291
\reference{8829} Berg, D. A., Skillman, E. D., Henry, R. B. C., Erb, D. K., \& Carigi, L. 
2016, ApJ, 827, 126
\reference{8822} Blanco-Cuaresma, S., Nordlander, T., Heiter, U., et al. 2016,  arXiv:  
1609.09071
\reference{1771} Bohlin, R. C. 1975, ApJ, 200, 402
\reference{2692} Boulanger, F., Prevot, M. L., \& Gry, C. 1994, A\&A, 284, 956
\reference{8816} Bromage, G. E., \& Nandy, K. 1983, MNRAS, 204, 29P
\reference{7335} Budzynski, J. M., \& Hewett, P. C. 2011, MNRAS, 416, 1871
\reference{8882} Burbidge, E. M., Burbidge, G. R., Fowler, W. A., \& Hoyle, F. 1957, 
RvMP, 29, 547
\reference{5096} Calura, F., Matteucci, F., \& Vladilo, G. 2003, MNRAS, 340, 59
\reference{8884} Cameron, A. G. W. 1957, PASP, 69, 201
\reference{5964} Cartledge, S. I. B., Clayton, G. C., Gordon, K. D., et al. 2005, ApJ, 630, 
355
\reference{5157} Cassinelli, J. P., Cohen, D. H., MacFarlane, J. J., et al. 1996, ApJ, 460, 
949
\reference{8772} Chen, H.-W., Kennicutt, R. C., Jr., \& Rauch, M. 2005, ApJ, 620, 703
\reference{5574} Chiappini, C., Matteucci, F., Beers, T. C., \& Nomoto, K. 1999, ApJ, 
515, 226
\reference{5719} Christensen, L., Schulte-Ladbeck, R. E., Sánchez, S. F., et al. 2005, 
A\&A, 429, 477
\reference{7345} Cooke, R., Pettini, M., Steidel, C. C., Rudie, G. C., \& Nissen, P. E. 
2011, MNRAS, 417, 1534
\reference{8103} Cox, N. L. J., Cordiner, M. A., Ehrenfreund, P., et al. 2007, A\&A, 
470, 941
\reference{8797} De Cia, A., Ledoux, C., Mattsson, L., et al. 2016, A\&A, 596, A97
\reference{8514} Den Hartog, E. A., Lawler, J. E., Sobeck, J. S., Sneden, C., \& Cowan, J. 
J. 2011, ApJS, 194, 35
\reference{5596} Dessauges-Zavadsky, M., Calura, F., Prochaska, J. X., D'Odorico, S., 
\& Matteucci, F. 2004, A\&A, 416, 79
\reference{6099} Dessauges-Zavadsky, M., Prochaska, J. X., D'Odorico, S., Calura, F., 
\& Matteucci, F. 2006, A\&A, 445, 93
\reference{6847} Dessauges-Zavadsky, M., Calura, F., Prochaska, J. X., D'Odorico, S., 
\& Matteucci, F. 2007, A\&A, 470, 431
\reference{1291} Draine, B. T., \& Lee, H. M. 1984, ApJ, 285, 89
\reference{5231} Draine, B. T. 2003, ARA\&A, 41, 241
\reference{8834} Draine, B. T., \& Li, A. 2007, ApJ, 657, 810
\reference{6464} Draine, B. T., Dale, D. A., Bendo, G., et al. 2007, ApJ, 663, 866
\reference{8764} Dufton, P. L., Ryans, R. S. I., Trundle, C., et al. 2005, A\&A, 434, 
1125
\reference{8843} Dufton, P. L., Ryans, R. S. I., Thompson, H. M. A., \& Street, R. A. 
2008, MNRAS, 385, 2261
\reference{8738} Dwek, E. 2016, ApJ, 825, 136
\reference{8771} Ellison, S. L., Kewley, L. J., \& Mallén-Ornelas, G. 2005, MNRAS, 
357, 354
\reference{8827} Engelbracht, C. W., Rieke, G. H., Gordon, K. D., et al. 2008, ApJ, 678, 
804
\reference{5931} Evans, I. N., \& Dopita, M. A. 1985, ApJS, 58, 125
\reference{5933} Evans, I. N. 1991, ApJS, 76, 985
\reference{8287} Fox, A., Richter, P., \& Fechner, C. 2014, A\&A, 572, A102
\reference{6040} Fox, A. J., Savage, B. D., \& Wakker, B. P. 2005, AJ, 130, 2418
\reference{8823} Gardiner, L. T., \& Noguchi, M. 1996, MNRAS, 278, 191
\reference{8818} Gordon, K. D., \& Clayton, G. C. 1998, ApJ, 500, 816
\reference{8819} Gordon, K. D., Clayton, G. C., Misselt, K. A., Landolt, A. U., \& Wolff, 
M. J. 2003, ApJ, 594, 279
\reference{8800} Guber, C. R., \& Richter, P. 2016, A\&A, 591, A137
\reference{6809} Habing, H. 1969, BAN, 20, 177
\reference{8855} Hagen, L. M. Z., Siegel, M. H., Hoversten, E. A., et al. 2016,  arXiv:  
1611.00064
\reference{8603} Hammer, F., Yang, Y. B., Flores, H., Puech, M., \& Fouquet, S. 2015, 
ApJ, 813, 110
\reference{8788} Hinkel, N. R., Young, P. A., Pagano, M. D., et al. 2016,  arXiv:  
1607.03130
\reference{4750} Hou, J. L., Boissier, S., \& Prantzos, N. 2001, A\&A, 370, 23
\reference{8084} Hunter, I., Dufton, P. L., Ryans, R. S. I., et al. 2005, A\&A, 436, 687
\reference{6375} Hunter, I., Dufton, P. L., Smartt, S. J., et al. 2007, A\&A, 466, 277
\reference{8757} Hunter, I., Brott, I., Langer, N., et al. 2009, A\&A, 496, 841
\reference{1540} Hutchings, J. B. 1982, ApJ, 255, 70
\reference{5589} Izotov, Y. I., \& Thuan, T. X. 1999, ApJ, 511, 639
\reference{8175} Jenkins, E. 2013, in The Life Cycle of Dust in the Universe: 
Observations, Theory, and Laboratory Experiments, eds. A. Anderssen, et al. (Taipei, 
Taiwan: Proceedings of Science), 15
\reference{1063} Jenkins, E. B., Savage, B. D., \& Spitzer, L. 1986, ApJ, 301, 355
\reference{3184} Jenkins, E. B. 1996, ApJ, 471, 292
\reference{6035} Jenkins, E. B., \& Tripp, T. M. 2006, ApJ, 637, 548
\reference{6999} Jenkins, E. B. 2009, ApJ, 700, 1299
\reference{8028} Kalberla, P. M. W., McClure-Griffiths, N. M., Pisano, D. J., et al. 2010, 
A\&A, 521, A17
\reference{7403} Khare, P., Vanden Berk, D., York, D. G., Lundgren, B., \& Kulkarni, 
V. P. 2012, MNRAS, 419, 1028
\reference{8007} Kisielius, R., Kulkarni, V. P., Ferland, G. J., Bogdanovich, P., \& 
Lykins, M. L. 2014, ApJ, 780, 76
\reference{8436} Kisielius, R., Kulkarni, V. P., Ferland, G. J., et al. 2015, ApJ, 804, 76
\reference{8770} Kobayashi, C., Umeda, H., Nomoto, K. i., Tominaga, N., \& Ohkubo, 
T. 2006, ApJ, 653, 1145
\reference{8066} Koenigsberger, G., Georgiev, L., Peimbert, M., et al. 2001, AJ, 121, 
267
\reference{8083} Korn, A. J., Becker, S. R., Gummersbach, C. A., \& Wolf, B. 2000, 
A\&A, 353, 655
\reference{4073} Kulkarni, V. P., Fall, S. M., \& Truran, J. W. 1997, ApJ, 484, L7
\reference{8842} Kulkarni, V. P., Som, D., Morrison, S., et al. 2015, ApJ, 815, 24
\reference{4901} Ledoux, C., Bergeron, J., \& Petitjean, P. 2002, A\&A, 385, 802
\reference{8765} Lee, J. K., Rolleston, W. R. J., Dufton, P. L., \& Ryans, R. S. I. 2005, 
A\&A, 429, 1025
\reference{4675} Lehner, N., Sembach, K. R., Dufton, P. L., Rolleston, W. R. J., \& 
Keenan, F. P. 2001, ApJ, 551, 781
\reference{6620} Lehner, N., Howk, J. C., Keenan, F. P., \& Smoker, J. V. 2008, ApJ, 
678, 219
\reference{7049} Levshakov, S. A., Agafonova, I. I., Molaro, P., Reimers, D., \& Hou, J. 
L. 2009, A\&A, 507, 209
\reference{8833} Li, A., \& Draine, B. T. 2002, ApJ, 576, 762
\reference{5604} Lodders, K. 2003, ApJ, 591, 1220
\reference{6037} Lopez, S., Reimers, D., Gregg, M. D., et al. 2005, ApJ, 626, 767
\reference{3446} Lu, L., Sargent, W. L. W., Barlow, T. A., Churchill, C. W., \& Vogt, S. S. 
1996, ApJS, 107, 475
\reference{8094} Luck, R. E., \& Lambert, D. L. 1992, ApJS, 79, 303
\reference{8752} Luck, R. E., Moffett, T. J., Barnes, T. G., III, \& Gieren, W. P. 1998, AJ, 
115, 605
\reference{8820} Maíz Apellániz, J., \& Rubio, M. 2012, A\&A, 541, A54
\reference{3866} Mallouris, C., Welty, D. E., York, D. G., et al. 2001, ApJ, 558, 133
\reference{5205} Mallouris, C. 2003, ApJS, 147, 265
\reference{5572} Matteucci, F. 2003, The Chemical Evolution of the Galaxy, Vol. 253 
(Dordrecht: Springer)
\reference{5818} McWilliam, A. 1997, ARA\&A, 35, 503
\reference{7004} Meiring, J. D., Lauroesch, J. T., Kulkarni, V. P., et al. 2009, MNRAS, 
397, 2037
\reference{8608} Morrison, S., Kulkarni, V. P., Som, D., et al. 2016, ApJ, 830, 158
\reference{1201} Morton, D. C. 1975, ApJ, 197, 85
\reference{5404} ---. 2003, ApJS, 149, 205
\reference{8065} Mucciarelli, A. 2014, AN, 335, 79
\reference{8879} Murphy, M. T., \& Bernet, M. L. 2016, MNRAS, 455, 1043
\reference{214} Pei, Y. C., Fall, S. M., \& Bechtold, J. 1991, ApJ, 378, 6
\reference{7154} Penprase, B. E., Prochaska, J. X., Sargent, W. L. W., Martinez, I. T., 
\& Beeler, D. J. 2010, ApJ, 721, 1
\reference{7429} Péroux, C., Bouché, N., Kulkarni, V. P., York, D. G., \& Vladilo, G. 
2011, MNRAS, 410, 2237
\reference{8773} Péroux, C., Bouché, N., Kulkarni, V. P., York, D. G., \& Vladilo, G. 
2012, MNRAS, 419, 3060
\reference{8774} Péroux, C., Kulkarni, V. P., \& York, D. G. 2014, MNRAS, 437, 3144
\reference{2665} Pettini, M., Smith, L. J., Hunstead, R. W., \& King, D. L. 1994, ApJ, 
426, 79
\reference{4063} Pettini, M., Smith, L. J., King, D. L., \& Hunstead, R. W. 1997, ApJ, 
486, 665
\reference{3635} Pettini, M., Ellison, S. L., Steidel, C. C., \& Bowen, D. V. 1999, ApJ, 
510, 576
\reference{8161} Pettini, M., Ellison, S. L., Steidel, C. C., Shapley, A. E., \& Bowen, D. 
V. 2000, ApJ, 532, 65
\reference{7331} Pettini, M., Zych, B. J., Steidel, C. C., \& Chaffee, F. H. 2008, MNRAS, 
385, 2011
\reference{3558} Press, W. H., Teukolsky, S. A., Vetterling, W. T., \& Flannery, B. P. 
2007, Numerical Recipes, The Art of Scientific Computing (3rd ed.; Cambridge: 
Cambridge Univ. Press)
\reference{8817} Prevot, M. L., Lequeux, J., Prevot, L., Maurice, E., \& Rocca-
Volmerange, B. 1984, A\&A, 132, 389
\reference{4481} Prochaska, J. X., \& Wolfe, A. M. 1999, ApJS, 121, 369
\reference{4796} ---. 2002, ApJ, 566, 68
\reference{5238} Prochaska, J. X., Howk, J. C., \& Wolfe, A. M. 2003, Natur, 423, 57
\reference{8799} Prochaska, J. X., O'Meara, J. M., Fumagalli, M., Bernstein, R. A., \& 
Burles, S. M. 2015, ApJS, 221, 2
\reference{8682} Quiret, S., Péroux, C., Zafar, T., et al. 2016, MNRAS, 458, 4074
\reference{7834} Rafelski, M., Wolfe, A. M., Prochaska, J. X., Neeleman, M., \& 
Mendez, A. J. 2012, ApJ, 755, 89
\reference{7991} Rafelski, M., Neeleman, M., Fumagalli, M., Wolfe, A. M., \& 
Prochaska, J. X. 2014, ApJ, 782, L29
\reference{6192} Rodríguez, E., Petitjean, P., Aracil, B., Ledoux, C., \& Srianand, R. 
2006, A\&A, 446, 791
\reference{8766} Rolleston, W. R. J., Dufton, P. L., McErlean, N. D., \& Venn, K. A. 
1999, A\&A, 348, 728
\reference{8080} Rolleston, W. R. J., Venn, K., Tolstoy, E., \& Dufton, P. L. 2003, 
A\&A, 400, 21
\reference{4110} Roth, K. C., \& Blades, J. C. 1997, ApJ, 474, L95
\reference{8072} Russell, S. C., \& Dopita, M. A. 1992, ApJ, 384, 508
\reference{8826} Sandstrom, K. M., Bolatto, A. D., Draine, B. T., Bot, C., \& 
Stanimirović, S. 2010, ApJ, 715, 701
\reference{1700} Sarazin, C. L. 1977, ApJ, 211, 772
\reference{110} Savage, B. D., \& Sembach, K. R. 1991, ApJ, 379, 245
\reference{328} ---. 1996, ARA\&A, 34, 279
\reference{8835} Savaglio, S. 2001, in The Extragalactic Infrared Background and its 
Cosmological Implications, eds. M. Harwit, \& M. G. Hauser (IAU Symp. 204), 307-
321
\reference{5942} Schulte-Ladbeck, R. E., König, B., Miller, C. J., et al. 2005, ApJ, 625, 
L79
\reference{181} Sembach, K. R., \& Savage, B. D. 1992, ApJS, 83, 147
\reference{6971} Sofia, U. J., Gordon, K. D., Clayton, G. C., et al. 2006, ApJ, 636, 753
\reference{7897} Som, D., Kulkarni, V. P., Meiring, J., et al. 2013, MNRAS, 435, 1469
\reference{8515} Som, D., Kulkarni, V. P., Meiring, J., et al. 2015, ApJ, 806, 25
\reference{8883} Suess, H. E., \& Urey, H. C. 1956, RvMP, 28, 53
\reference{8881} Sukhbold, T., Ertl, T., Woosley, S. E., Brown, J. M., \& Janka, H. T. 
2016, ApJ, 821, 38
\reference{8449} Tchernyshyov, K., Meixner, M., Seale, J., et al. 2015, ApJ, 811, 78
\reference{5616} Thuan, T. X., Izotov, Y. I., \& Lipovetsky, V. A. 1995, ApJ, 445, 108
\reference{2991} Timmes, F. X., Woosley, S. E., \& Weaver, T. A. 1995, ApJS, 98, 617
\reference{7791} Tolstoy, E., Hill, V., \& Tosi, M. 2009, ARA\&A, 47, 371
\reference{8768} Trundle, C., Lennon, D. J., Puls, J., \& Dufton, P. L. 2004, A\&A, 417, 
217
\reference{8756} Trundle, C., Dufton, P. L., Hunter, I., et al. 2007, A\&A, 471, 625
\reference{3886} Tumlinson, J., Shull, J. M., Rachford, B. L., et al. 2002, ApJ, 566, 857
\reference{298} Vallerga, J. V., \& Welsh, B. Y. 1995, ApJ, 444, 702
\reference{8073} Venn, K. A. 1999, ApJ, 518, 405
\reference{5425} Vladilo, G., Centurión, M., Bonifacio, P., \& Howk, J. C. 2001, ApJ, 
557, 1007
\reference{4905} Vladilo, G. 2002, ApJ, 569, 295
\reference{5519} Vladilo, G. 2004, A\&A, 421, 479
\reference{2807} Wallerstein, G., \& Goldsmith, D. 1974, ApJ, 187, 237
\reference{4810} Weingartner, J. C., \& Draine, B. T. 1999, ApJ, 517, 292
\reference{8825} Weingartner, J. C., \& Draine, B. T. 2001, ApJ, 548, 296
\reference{4046} Welty, D. E., Lauroesch, J. T., Blades, J. C., Hobbs, L. M., \& York, D. 
G. 1997, ApJ, 489, 672
\reference{5126} ---. 2001, ApJ, 554, L75
\reference{7318} Welty, D. E., Federman, S. R., Gredel, R., Thorburn, J. A., \& 
Lambert, D. L. 2006, ApJS, 165, 138
\reference{7110} Welty, D. E., \& Crowther, P. A. 2010, MNRAS, 404, 1321
\reference{7456} Welty, D. E., Xue, R., \& Wong, T. 2012, ApJ, 745, 173
\reference{7733} Welty, D. E., Howk, J. C., Lehner, N., \& Black, J. H. 2013, MNRAS, 
428, 1107
\reference{8722} Welty, D. E., Lauroesch, J. T., Wong, T., \& York, D. G. 2016, ApJ, 
821, 118
\reference{2764} Wheeler, J. C., Sneden, C., \& Truran, J. W., Jr. 1989, ARA\&A, 27, 
279
\reference{8832} Wiseman, P., Schady, P., Bolmer, J., et al. 2016,  arXiv:  1607.00288
\reference{6064} Wolfe, A. M., Gawiser, E., \& Prochaska, J. X. 2005, ARA\&A, 43, 
861
\reference{8880} Woosley, S. E., Heger, A., \& Weaver, T. 2002, RvMP, 74, 1015
\reference{8828} Wu, Y., Charmandaris, V., Hao, L., et al. 2006, ApJ, 639, 157
\reference{6158} York, D. G., Khare, P., Vanden Berk, D., et al. 2006, MNRAS, 367, 
945
\reference{8821} Zonca, A., Casu, S., Mulas, G., Aresu, G., \& Cecchi-Pestellini, C. 
2015, ApJ, 810, 70
\reference{5477} Zubko, V., Dwek, E., \& Arendt, R. G. 2004, ApJS, 152, 211
\end{references}
\end{document}